\documentclass[a4paper,fleqn,usenatbib]{mnras}

\usepackage{mathptmx}

\usepackage[T1]{fontenc}
\usepackage{ae,aecompl}

\usepackage{graphics,graphicx,wrapfig}	
\usepackage{amsmath,caption}	
\usepackage{amssymb}	
\usepackage{epsfig,xcolor}
\usepackage{hyperref,tikz}

\def\kms{\,km\,s$^{-1}$}
\def\msun{\,M$_{\odot}$}

\newcommand*{\reffig}[1]{Fig.\,\ref{#1}}
\newcommand*{\refeq}[1]{Eq.\,\ref{#1}}
\newcommand*{\reftable}[1]{Table\,\ref{#1}}
\newcommand*{\refsec}[1]{Section\,\ref{#1}}

\title[WISDOM project -- V. NGC\,0383]{WISDOM project -- V. Resolving molecular gas in Keplerian rotation around the supermassive black hole in NGC\,0383}

\author[E. V. North et al.]{\parbox{\textwidth}{
	Eve V. North,$^{1}$\thanks{E-mail: NorthEV@cardiff.ac.uk}
	Timothy A. Davis,$^{1}$
	Martin Bureau,$^{2,3}$
	Michele Cappellari,$^{2}$
	Satoru Iguchi,$^{4,5}$
	Lijie Liu,$^{2}$
	Kyoko Onishi,$^{6,7}$
	Marc Sarzi,$^{8}$
	Mark D. Smith$^{2}$
	and Thomas G. Williams$^{1}$ }
    \vspace{0.4cm}
	\\
	\parbox{\textwidth}{
	$^{1}$School of Physics \& Astronomy, Cardiff University, Queens Buildings, The Parade, Cardiff CF24 3AA, UK\\
	$^{2}$Sub-department of Astrophysics, Department of Physics, University of Oxford, Denys Wilkinson Building, Keble Road, Oxford OX1 3RH, UK\\
	$^{3}$Yonsei Frontier Lab and Department of Astronomy, Yonsei University, 50 Yonsei-ro, Seodaemun-gu, Seoul 03722, Republic of Korea\\
	$^{4}$Department of Astronomical Science, SOKENDAI (The Graduate University of Advanced Studies), Mitaka, Tokyo 181-8588, Japan\\
	$^{5}$National Astronomical Observatory of Japan, National Institutes of Natural Sciences, Mitaka, Tokyo 181-8588, Japan\\
	$^{6}$Research Center for Space and Cosmic Evolution, Ehime University, Matsuyama, Ehime, 790-8577, Japan\\
	$^{7}$Department of Space, Earth and Environment, Chalmers University of Technology, Onsala Observatory, SE-439 92 Onsala, Sweden
	$^{8}$Armagh Observatory and Planetarium, College Hill, Armagh, BT61 9DG, UK\\}
}

\date{Accepted XXX. Received YYY; in original form ZZZ}

\pubyear{2019}

\begin{document}
\label{firstpage}
\pagerange{\pageref{firstpage}--\pageref{lastpage}}
\maketitle

\begin{abstract}
As part of the mm-Wave Interferometric Survey of Dark Object Masses (WISDOM), we present a measurement of the mass of the supermassive black hole (SMBH) in the nearby early-type galaxy NGC\,0383 (radio source 3C\,031). This measurement is based on Atacama Large Millimeter/sub-milimeter Array (ALMA) cycle 4 and 5 observations of the $^{12}$CO(2--1) emission line with a spatial resolution of  $58\times32$\,pc$^{2}$ ($0\farcs18\times0\farcs1$). 
This resolution, combined with a channel width of 10\kms, allows us to well resolve the radius of the black hole sphere of influence {(measured as $R_{\rm SOI}=316$\,pc\,=\,0\farcs98)}, where we detect a clear Keplerian increase of the rotation velocities. NGC\,0383 has a kinematically-relaxed, smooth nuclear molecular gas disc with weak ring/spiral features. 
We forward-model the ALMA data cube with the \textsc{Kinematic Molecular Simulation} (KinMS) tool and a Bayesian Markov Chain Monte Carlo method to measure a SMBH mass of $(4.2\pm0.7)\times10^{9}$\msun, a F160W-band stellar mass-to-light ratio that varies from $2.8\pm0.6$\msun/L$_{\odot,\,\mathrm{F160W}}$ in the centre to $2.4\pm0.3$\msun$/\rm L_{\odot,\,\mathrm{F160W}}$ at the outer edge of the disc and a molecular gas velocity dispersion of $8.3\pm2.1$\kms (all $3\sigma$ uncertainties). 
We also detect unresolved continuum emission across the full bandwidth, consistent with synchrotron emission from an active galactic nucleus.
This work demonstrates that low-J CO emission can resolve gas very close to the SMBH ($\approx140\,000$ Schwarzschild radii) and hence that the molecular gas method is highly complimentary to megamaser observations as it can probe the same emitting material.
\end{abstract}




\section{Introduction}
\label{sec:intro}


Early-type galaxies, as gravitationally-bound stellar systems, lie on a tight ``Fundamental Plane'' defined by their mass (luminosity), size (half-light radius) and second velocity moment (velocity dispersion; e.g. \citealt{Djorgovski1987,Dressler1987}). Late-type galaxies follow less tight correlations such as the Tully-Fisher relation between mass (luminosity) and rotation velocity (\citealt{Tully1977}; see \citealt{Courteau2014} and Section 4 of \citealt{Cappellari2016} for reviews of the fundamental planes of galaxies).
Comparing central supermassive black hole (SMBH) mass measurements with these galaxy properties has revealed further relations connecting, for example, bulge mass, stellar mass (or luminosity) or Sersic concentration index to the SMBH mass (\citealt{Kormendy1995}; \citealt{Magorrian1998};\\ \citealt{Marconi2003,Haering2004,Graham2007}).
This has led to the prevailing theory that SMBHs, despite their comparatively small masses, are a major influence on galaxy evolution \citep[e.g.][]{Beifiori2012,Bosch2016}.
However, these relations are poorly constrained, with relatively few data points drawn from biased samples, and with large uncertainties.
{Further evidence indicates both SMBH mass growth at the same rate as \citep[e.g.][]{Mullaney2012,MadauDickinson2014} and SMBH feedback quenching of \citep{Bundy2008} star formation by SMBHs.}
Furthermore, whether all galaxies follow the same relations or not is still inadequately tested.
In particular, there is evidence that low-mass late-type and high-mass early-type galaxies follow different co-evolutionary relationships \citep[e.g.][]{McConnell2013}. \citet{Kormendy2013} give a comprehensive review of the current state of these relations.

One of the tightest relationships is that between the SMBH mass ($M_{\mathrm{BH}}$) and the stellar velocity dispersion ($\sigma_{\ast}$; e.g. \citealt{Gebhardt2000,Ferrarese2000}), 
but there is again growing evidence of divergence between galaxies of different morphological types or masses (e.g. \citealt{McConnell2013,Bosch2016}, esp. their Fig. 2; \citealt{Krajnovic2018}). 
To fully analyse the extent of the co-evolution between all these galaxy properties, it is essential to gather a larger, more diverse, sample of \emph{reliable} SMBH mass estimates \citep{Bosch2016}. 

Reliability is achieved by directly measuring the SMBH masses through their gravitational influence.
Methods to measure SMBH masses dynamically include observing and modelling the stellar kinematics \citep[e.g.][]{Dressler1988,Cappellari2002,Krajnovi2009}, ionised gas kinematics  \citep[e.g.][]{Ferrarese1996,Sarzi2001,Walsh2013} and megamaser kinematics \citep[e.g.][]{Herrnstein1999,Miyoshi1995,Greene2010}.
However, each of these methods can only be used in a small fraction of the galaxy population, as each is biased towards particular morphologies.
For instance, stellar kinematics are often hampered by dust contamination and require either resolving individual stars directly or strong absorption lines in integrated spectra.
Megamasers probe material very close to the SMBHs but require an edge-on view and are very rare {(being present in only $\approx5\,$\% of objects searched; \citealt{Lo2005}). They are typically found in Seyfert 2 and low-ionisation nuclear emission region (LINER)-type nuclei of low-mass galaxies.}  
Overall the current sample is biased towards nearby, high surface brightness objects. A new method of measuring SMBH masses is thus required to diversify the sample.

To expand the current sample to all morphological types, galaxy masses and both active and non-active galaxies, our mm-Wave Interferometric Survey of Dark Object Masses (WISDOM) is using a new method exploiting molecular gas observations to trace the velocity fields surrounding SMBHs. The first use of this method with Carbon Monoxide (CO) was by \citet{Davis2013b}. SMBH mass measurements in fast-rotator early-type galaxies \citep{Onishi2017,Davis2017,Davis2018}, galaxies with irregular gas distributions \citep{Smith2019}, and in the first late-type galaxy with the dense molecular gas tracers HCN and HCO$^{+}$ \citep{Onishi2015} have been successful.
\citet{Barth2016a, Barth2016b} and \citet{Boizelle2019} also used CO to constrain the SMBH mass in the early-type galaxy NGC\,1332 and NGC\,3258. Most recently, \citet{Combes2019} used CO(3--2) observations to investigate the molecular tori around 7 SMBHs and therefore measure their SMBH masses. \citet{Nagai2019} observed the radial filaments of NGC\,1275 in CO(2--1), detecting a rotating disc allowing them to make a SMBH mass estimate, {that agrees with the estimate from H$_{2}$ observations by \citet{Scharwachter2013}}.

All these observations can detect the dynamical influence of the SMBH if, as shown in \citet{Davis2014}, they have a minimum spatial resolution of approximately two times the radius of the sphere of influence ($R_{\mathrm{SOI}}$) of the SMBH,
\begin{equation}
{R_{\mathrm{SOI}}} \equiv \frac{GM_{\mathrm{BH}}}{\sigma^{2}_{\ast}}\,,
\label{eq:SOI}
\end{equation}
where $G$ is the gravitational constant. 
The use of molecular gas, specifically $^{12}$CO, reduces the selection biases normally associated with dynamical SMBH mass measurements, because of the wide range of objects with suitable molecular gas discs, and because the high angular resolution required is easily reached by modern interferometers, e.g. the Atacama Large Millimeter/sub-millimeter Array (ALMA).
Indeed, molecular discs are found around the centres of galaxies of all morphological types \citep[e.g.][]{Regan2001,Alatalo2013}. 
Furthermore, with rotational transitions in the millimetre/sub-millimetre wavebands, CO is observable without dust attenuation. 

{NGC\,0383 (radio source 3C\,031; \citealt{Edge1959,Bennett1962}) is a well-known radio galaxy (implying the presence of a large SMBH), it has a very regular central dust disc and it is also strongly detected in CO with a clear double-horned profile \citep{Lim2000,Okuda2005,OcanaFlaquer2010}.
This work presents a measurement of the SMBH mass in this galaxy using ALMA observations of the $^{12}$CO(2--1) line with a spatial resolution of $58\times32$\,pc$^{2}$ ($0\farcs18\times0\farcs1$).} 
In \refsec{sec:Obs}, we present the target, observations and data reduction. In \refsec{sec:model}, we describe the dynamical modelling and SMBH mass measurement method. A discussion of the uncertainties and comparisons to other SMBH mass measurement techniques are presented in \refsec{sec:discussion}. We conclude briefly in \refsec{sec:conclusion}.

\section{Target: NGC 0383}
\label{sec:Obs}

NGC\,0383 is a dusty lenticular galaxy at a distance of $66.6\pm9.9$\,Mpc \citep{Freedman2001}. It is the brightest galaxy of its group (the NGC\,0383 group), part of the Pisces-Perseus Supercluster \citep{Hudson2001}. 
NGC\,0383 hosts a radio-loud active galactic nucleus (AGN) with spectacular radio jets. The coincident radio source is catalogued as 3C\,031 \citep{Edge1959,Bennett1962}.
Observations of the jets are presented in \cite{Macdonald1968}, \cite{Bridle1984}, \cite{Laing2002} and \cite{VanVelzen2012}, while the flat-spectrum radio source is described in \citet{Healey2007}.

We can estimate the required angular resolution by predicting the $R_{\mathrm{SOI}}$ using the SMBH mass upper limit of \citet{Beifiori2009} ($M_{\mathrm{BH}}=1.1\times10^{9}$\msun, corrected to our assumed distance and inclination, see \refsec{sec:lit}) and $\sigma_{\ast}$ as listed in \citet{Bosch2016} ($\sigma_{\rm e}=239\pm16$\,\kms\ i.e. $\sigma_{\ast}$ within $R_{\mathrm{e}}$ from \citealt{Beifiori2009}, corrected following \citealt{Jorgensen1995}). Using \refeq{eq:SOI} with these values, we obtain $R_{\mathrm{SOI}}=82\pm15$\,pc, indicating we need an angular resolution of better than $\approx0$\farcs5 to attempt to detect the dynamical influence of the SMBH (i.e. to resolve $2\,R_{\mathrm{SOI}}$; \citealt{Davis2014}).

{There are existing $^{12}$CO(1--0) observations of NGC\,0383 from \citet{Lim2000}, \citet{Okuda2005} and in particular single dish observations from the Thorough ANalysis of radio-Galaxies Observation project (TANGO; \citealt{OcanaFlaquer2010}). They report the total molecular gas mass enclosed by the Institut de Radioastronomie Millim\'{e}trique (IRAM) 30-m telescope beam to be $M_{\rm H_{2}}=(1.7\pm0.2)\times10^{9}$\msun. We correct this from $71.06\,$Mpc to our assumed distance of $66.6\,$Mpc, yielding $(1.49\pm0.19)\times10^{9}$\msun\ as the total molecular gas mass.}

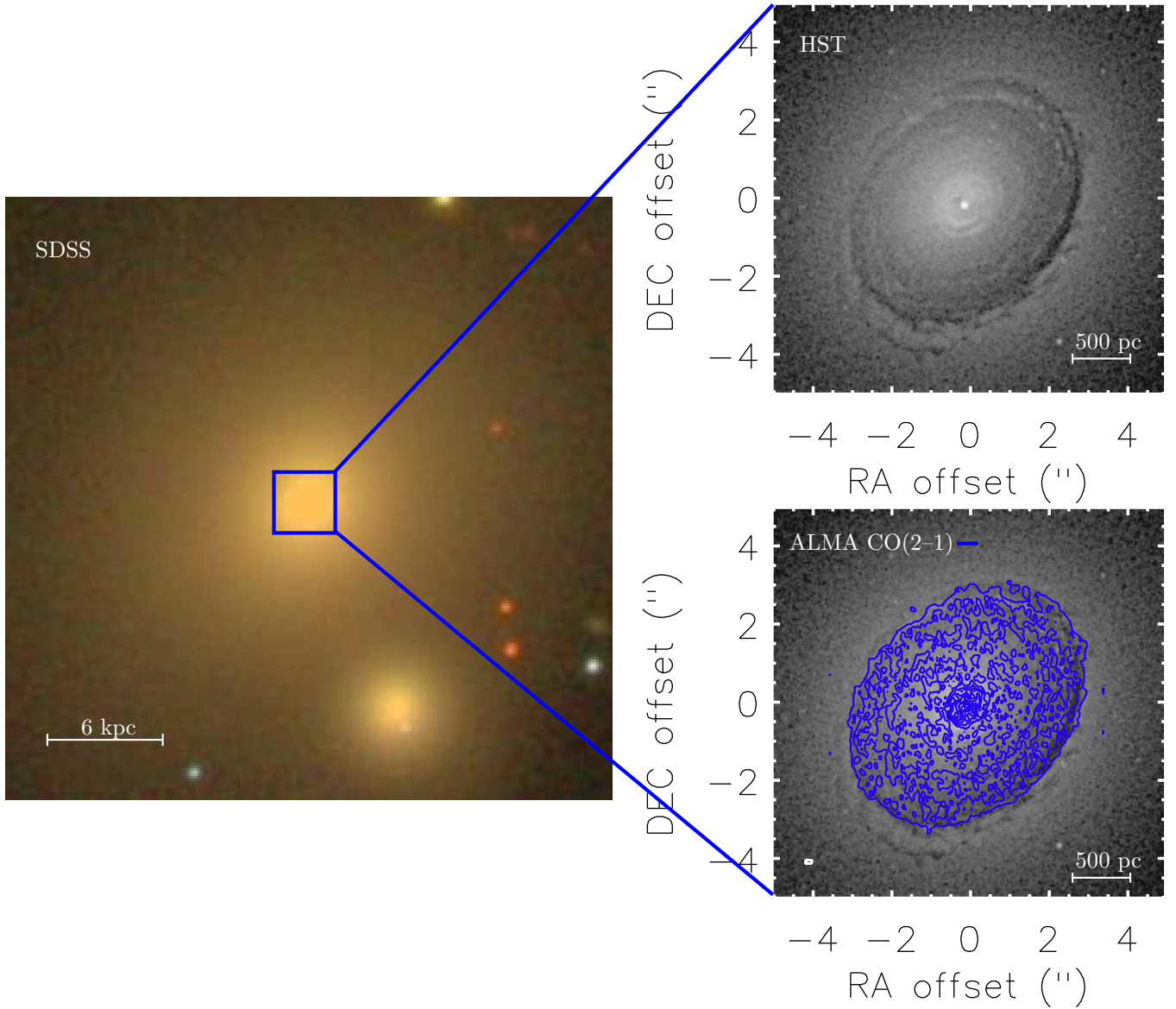
\begin{figure*}
	\begin{center}
		\begin{tikzpicture}
		\node[anchor=south west,inner sep=0] (image) at (0,0) {
		\begin{minipage}[b]{0.52\textwidth}	
		\begin{tikzpicture}
			\node[anchor=south west,inner sep=0] (image) at (0,0) {\includegraphics[height=9.25cm,angle=0,clip,trim=0cm 0cm 0cm 0.0cm]{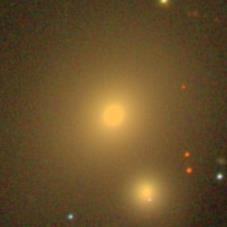}};
			\begin{scope}[x={(image.south east)},y={(image.north west)}]
			\draw[blue,ultra thick] (0.443,0.443) rectangle (0.544,0.544);
			\node[text=white] at (0.05,0.9) {\large SDSS};
			\draw[white,thick] (0.07,0.1) -- (0.260,0.1);	
			\draw[white,thick] (0.07,0.09) -- (0.07,0.11);	
			\draw[white,thick] (0.260,0.09) -- (0.260,0.11);	
			\node[text=white] at (0.125,0.10) {\large 6 kpc};
			\end{scope}
		\end{tikzpicture}\vspace{3.2cm}
		\end{minipage}\hspace{0.5cm}
		\begin{minipage}[b]{0.45\textwidth}
		\begin{tikzpicture}
			\node[anchor=south west,inner sep=0] (image) at (0,0) {\includegraphics[width=8cm,angle=0,clip,trim=0cm 0cm 0cm 0.0cm]{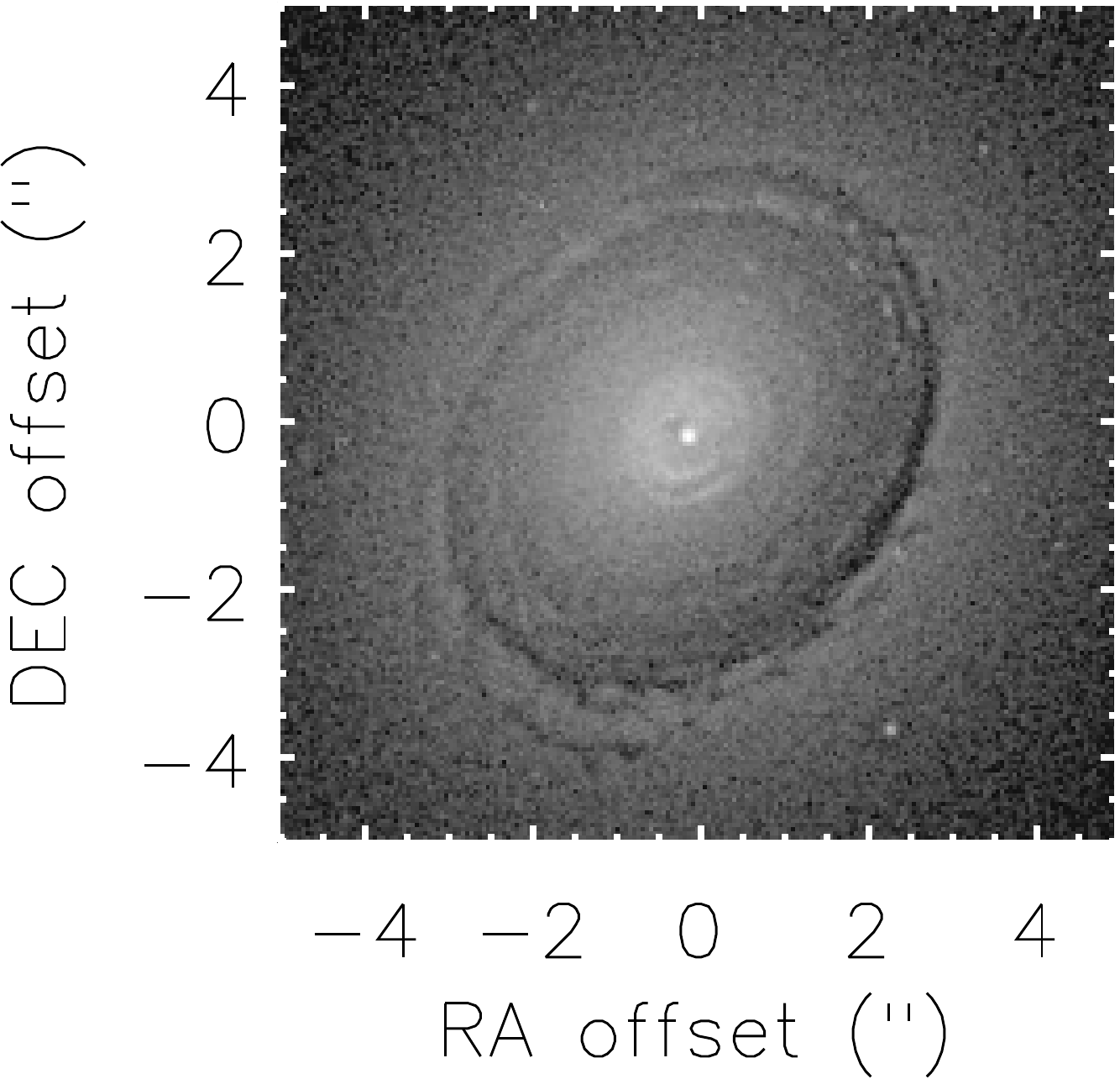}};
			\begin{scope}[x={(image.south east)},y={(image.north west)}]
			\draw[white,thick] (0.82,0.29) -- (0.93,0.29);	
			\draw[white,thick] (0.82,0.28) -- (0.82,0.30);	
			\draw[white,thick] (0.93,0.28) -- (0.93,0.30);	
			\node[text=white] at (0.825,0.30) {\large 500 pc};
			\node[text=white] at (0.3,0.9) {\large HST};
			\end{scope}
		\end{tikzpicture}
		\begin{tikzpicture}
			\node[anchor=south west,inner sep=0] (image) at (0,0) {\includegraphics[width=8cm,angle=0,clip,trim=0cm 0cm 0cm 0.0cm]{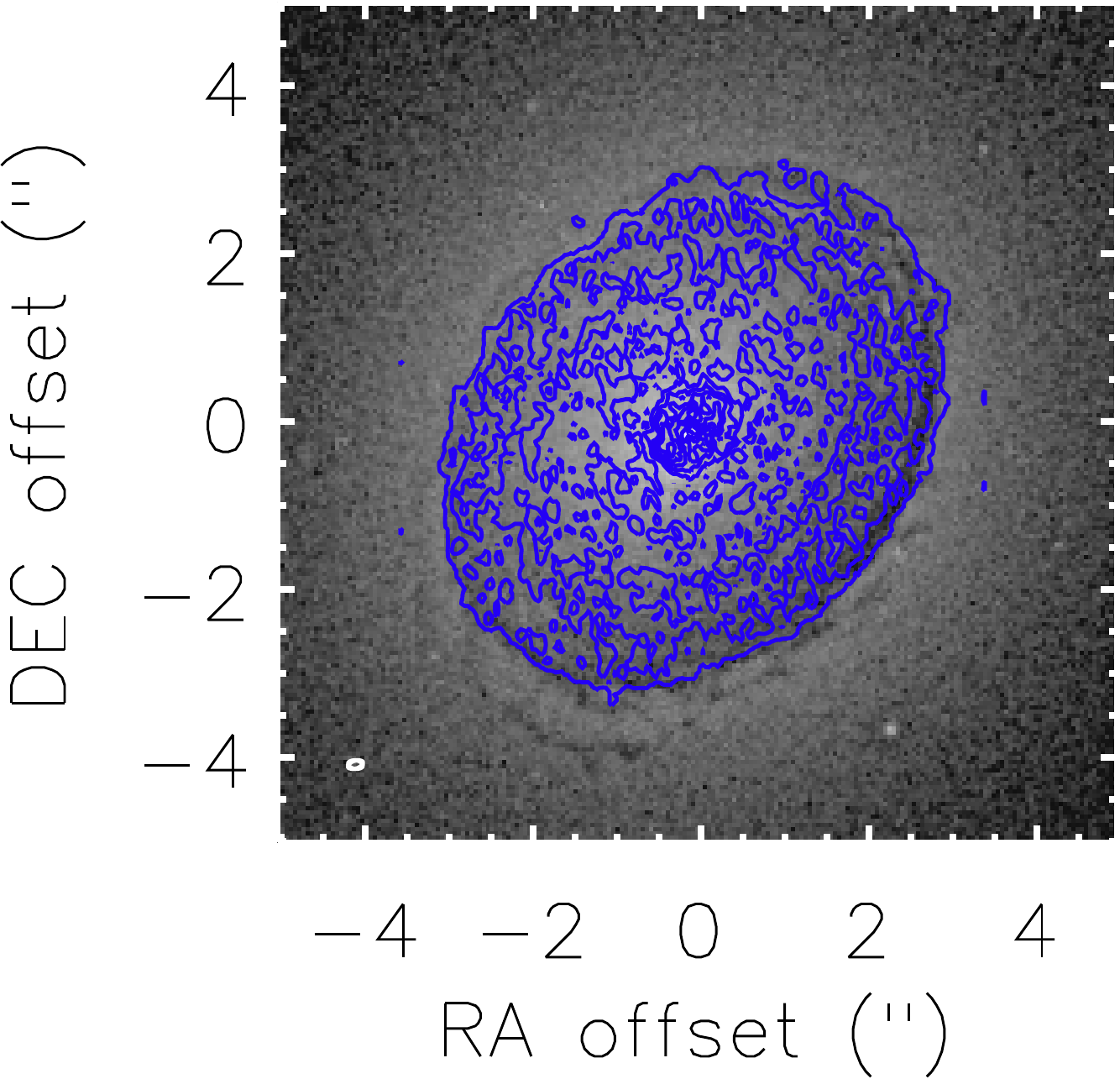}};
			\begin{scope}[x={(image.south east)},y={(image.north west)}]
			\draw[white,thick] (0.82,0.26) -- (0.93,0.26);	
			\draw[white,thick] (0.82,0.25) -- (0.82,0.27);	
			\draw[white,thick] (0.93,0.25) -- (0.93,0.27);	
			\node[text=white] at (0.825,0.27) {\large 500 pc};
			\node[text=white] at (0.28,0.9) {\large ALMA CO(2--1)};
			\draw[blue,ultra thick] (0.6,0.925) -- (0.64,0.925);	
			\end{scope}
		\end{tikzpicture}
	\end{minipage}
};
	\begin{scope}[x={(image.south east)},y={(image.north west)}]
		\draw[blue,ultra thick] (0.285,0.5338) -- (0.662,0.997);	
		\draw[blue,ultra thick] (0.285,0.4737) -- (0.662,0.113);
	\end{scope}
\end{tikzpicture}
\caption{{\textit{Left panel:} SDSS three-colour ($gri$) image of NGC\,0383, $90\arcsec\times90\arcsec$ (29\,kpc $\times$ 29\,kpc) in size. \textit{Right panel, top:} Unsharp-masked Hubble Space telescope (\textit{HST}) Wide-Field Planetary Camera 2 (WFPC2) F555W image of a 3.2\,kpc\,$\times$\,3.2\,kpc region around the nucleus (indicated in blue in the left panel), revealing a clear central dust disc. \textit{Right panel, bottom:} As above, but overlaid with blue $^{12}$CO(2--1) integrated intensity contours from our ALMA observations. The synthesised beam ($0\farcs18\,\times\,0\farcs1$ or 58\,$\times$\,32 pc$^2$) is shown as a (very small) white ellipse in the bottom-left corner. The molecular gas disc coincides with the dust disc.}}
\label{gal_overview}
\end{center}
\end{figure*}

\subsection{ALMA Observations}
\label{sec:ALMA}

The  $^{12}$CO(2--1) line in NGC\,0383 was observed with ALMA on the 21st of June 2016 at moderate resolution (0\farcs5) and then on the 16th of August 2017 at high resolution (0\farcs1), both as part of the WISDOM project (programmes 2015.1.00419.S and 2016.1.00437.S). 
Configurations C36-5 (baselines 15--704\,m) and C40-8 (baselines 21--3637\,m) were used to achieve sensitivity to emission on scales up to 4\arcsec, with on-source integration times of 2.22 and 28.8\,min, respectively. 
A 1850\,MHz correlator window was placed over the CO(2--1) line {and centred at 226.6\,GHz}, yielding a continuous velocity coverage of $\approx2000$\kms\ with a raw channel width of $\approx1.3$\kms, fully covering and well resolving the line. Three additional low spectral resolution correlator windows were included to detect continuum emission, each of 2\,GHz width. 
 
The raw ALMA data were calibrated using the standard ALMA pipeline, as provided by the ALMA regional centre staff. The amplitude and bandpass calibrator used in the two observations was respectively J0237+2848 and J2253+1608. 
The phase calibration used J0057+3021 and J0112+3208, respectively, to determine and therefore correct atmospheric phase offsets.

We then used the \textsc{Common Astronomy Software Applications} (\textsc{CASA}; \citealt{McMullin2007}) package to combine the two configurations and image the resultant visibilities. 
A three-dimensional RA-Dec-velocity data cube was produced with a binned channel width of 10\kms. 
To balance spatial sampling and resolution, pixels of 0\farcs035$\times$0\farcs035 were chosen, yielding approximately 5 pixels across the synthesised beam major axis.  
 
The data presented here were produced using Briggs weighting with a robust parameter of 0.5, yielding a synthesised beam full-width-at-half-maximum (FWHM) of $\theta_{\mathrm{maj}}\times\theta_{\mathrm{min}}\approx0\farcs18\times0\farcs1$ at a position angle of 6\fdg6.
The corresponding spatial resolution is $\approx58\times32$\,pc$^{2}$, so that the predicted $R_{\mathrm{SOI}}$ is well resolved with about 2 synthesised beams, i.e. $R_{\mathrm{SOI}}\,/\sqrt{\theta_{\mathrm{maj}}\times\theta_{\mathrm{min}}} = 1.9$.
Continuum emission was detected, measured over the full line-free bandwidth, and then subtracted from the data in the $uv$--plane using the \textsc{CASA} task \textsc{uvcontsub}. 
The achieved continuum root mean square (RMS) noise is $35\,\mu$Jy\,beam$^{-1}$. 
The continuum-subtracted dirty cube was cleaned (see \citealt{Hogbom1969} for the \textsc{CLEAN} procedure) in interactively-identified regions of source emission in each channel, to a threshold equal to the RMS noise of the dirty channels.
The clean components were then added back and re-convolved using a Gaussian beam of FWHM equal to that of the dirty beam.
This produced the final, reduced and fully calibrated $^{12}$CO(2--1) data cube of NGC0383, with a RMS noise level of 0.4\,mJy\,beam$^{-1}$ in each 10\kms\ channel.

\begin{figure*}
	\includegraphics[height=.33\textwidth,trim=0cm 0cm 0cm 0cm,clip]{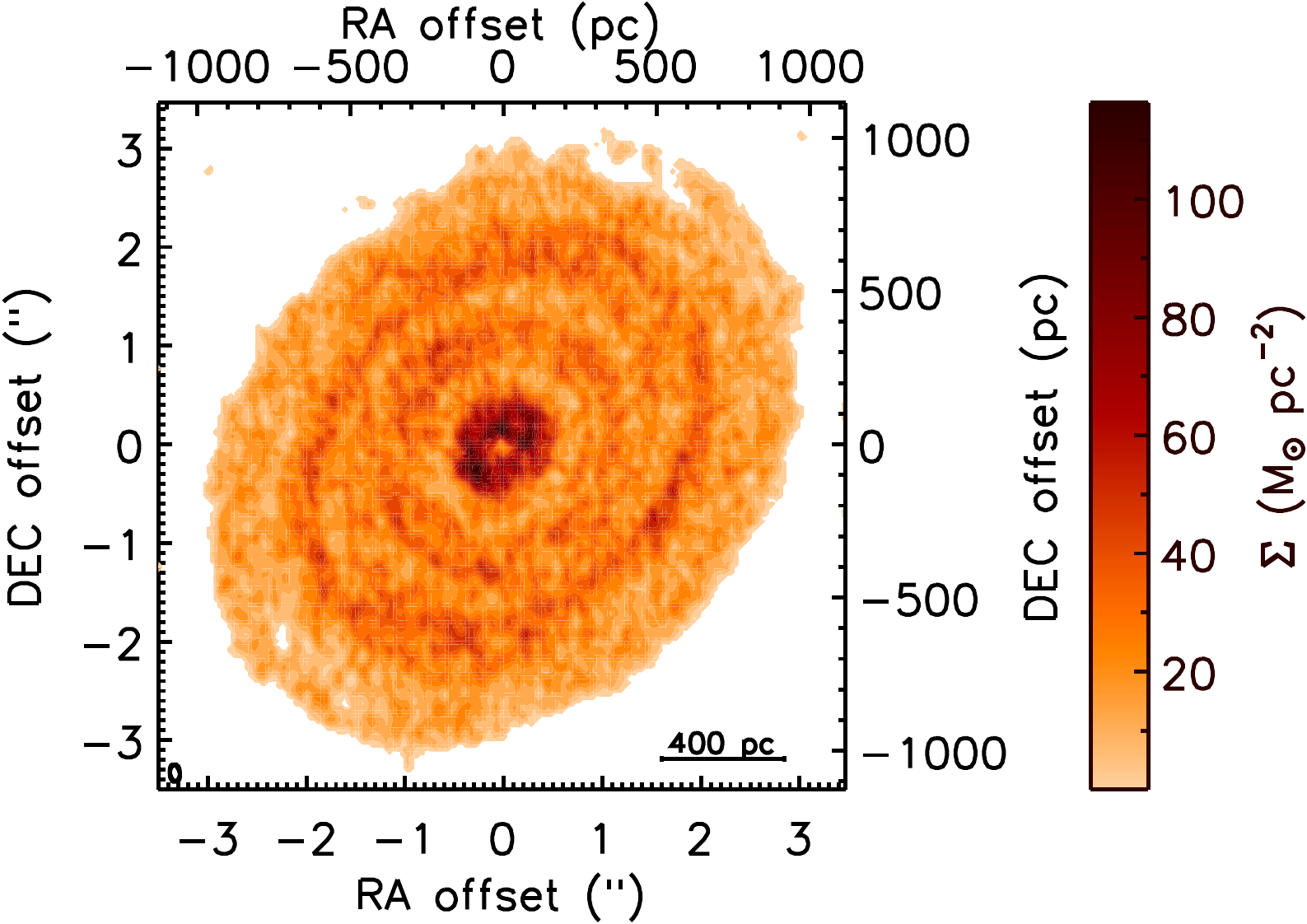}
	\includegraphics[height=.33\textwidth,trim=86cm 0cm 0cm 0cm,clip]{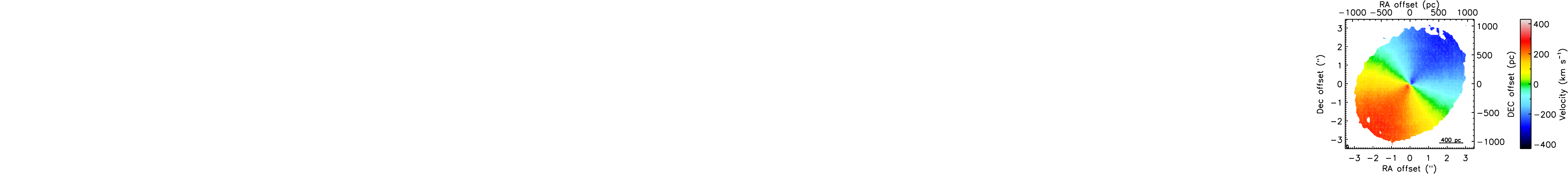}
	\caption{{Moment zero (integrated intensity; left-hand panel) map of NGC\,0383 assuming conversion factor CO-to-H$_{2}$ $\alpha_{\rm CO}=4.8$\msun(K \kms)$^{-1}$ pc$^{-2}$. Moment one (intensity-weighted mean velocity; right-hand panel) of NGC\,0383. The ellipse at the bottom-left of each panel shows the synthesised beam (0\farcs18$\times$0\farcs1).}}
	\label{fig:mom01}
\end{figure*}

\begin{figure}
	\includegraphics[width=\columnwidth]{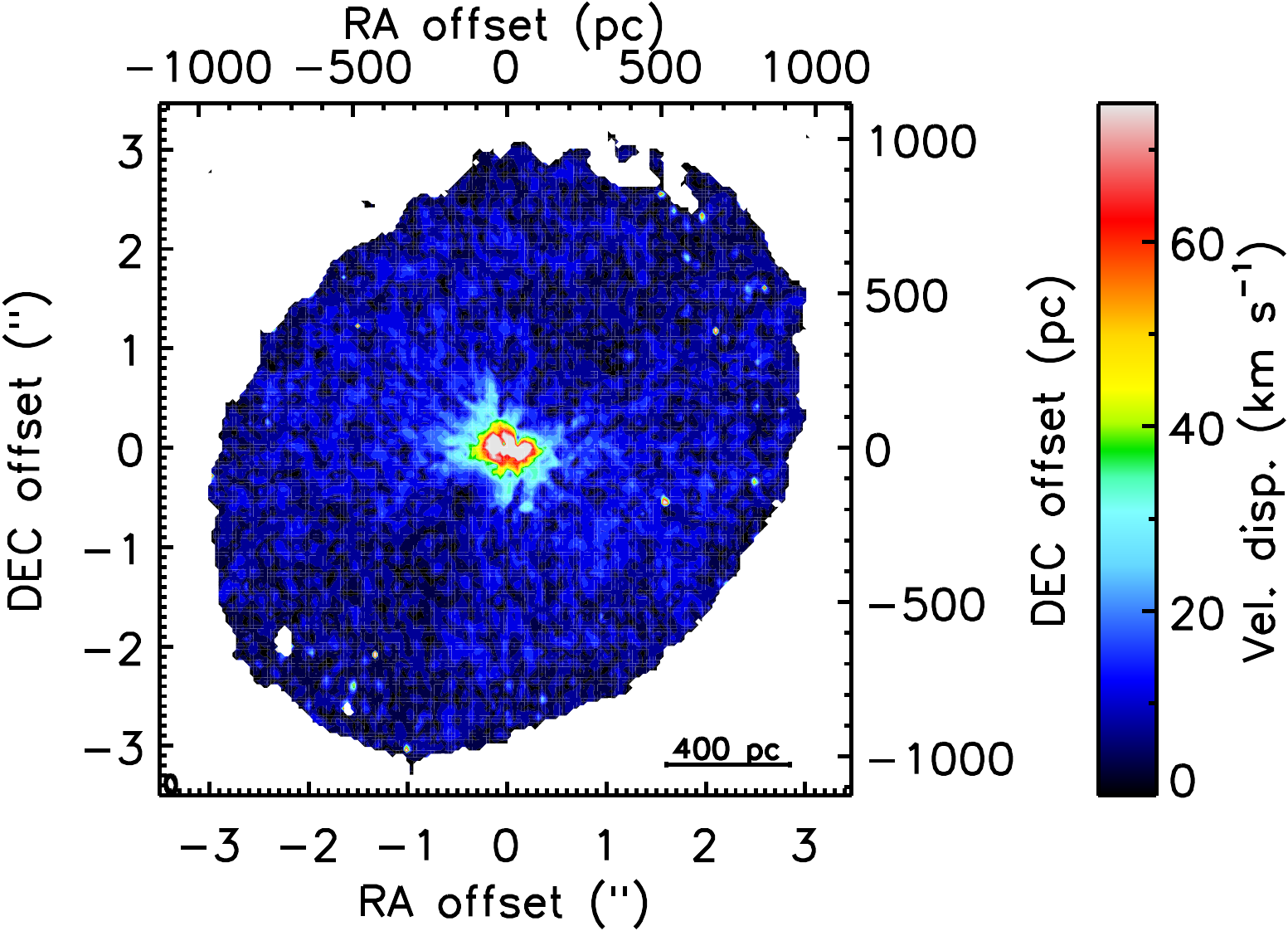}
	\caption{Moment two (intensity-weighted velocity dispersion) map of NGC\,0383. The ellipse at the bottom-left shows the synthesised beam (0\farcs18$\times$0\farcs1).}
	\label{fig:mom2}
\end{figure}


\subsection{Line Emission}
\label{sec:line}

The final data products used in this paper were created from the clean, fully calibrated data cube. Zeroth moment (integrated intensity), first moment (mean velocity) and second moment (velocity dispersion) maps were created using a masked moment technique \citep[e.g.][]{Dame2011}. The mask was generated by taking a copy of the clean cube and smoothing it, first spatially using a Gaussian with FWHM equal to that of the synthesised beam, and then Hanning-smoothing in velocity. 
The mask selects pixels with an amplitude in the smoothed cube greater than 0.8 times the RMS of the unsmoothed data cube. The moments, shown in Figs. \ref{fig:mom01} and \ref{fig:mom2}, are made from the original un-smoothed cube with the mask applied. We note that the masking procedure is only used when creating the moment maps, whilst the fitting is performed on the whole unmasked cube.

\begin{figure}
	\includegraphics[width=\columnwidth]{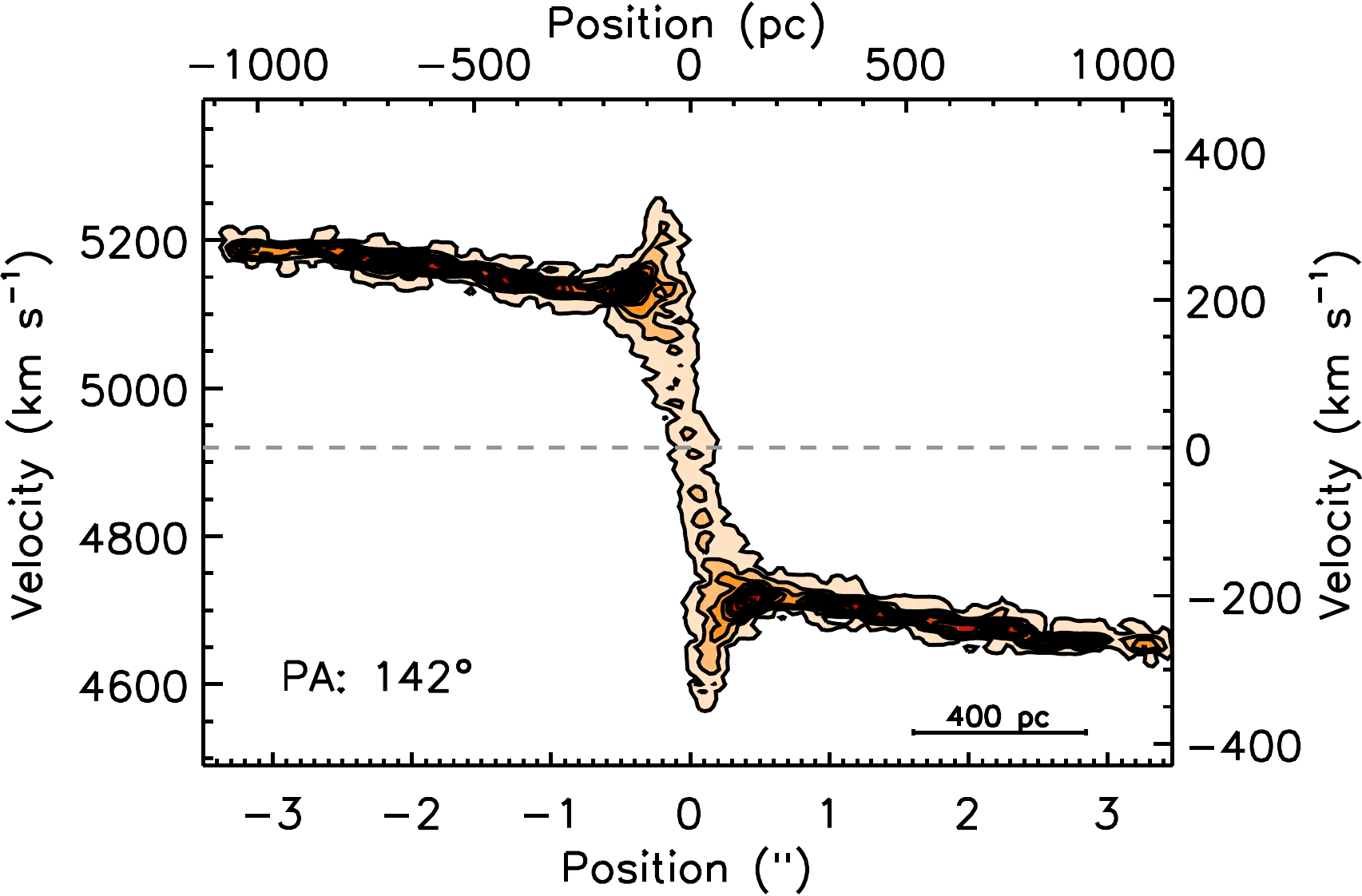}
	\caption{Major-axis position-velocity diagram of NGC\,0383. The SMBH signature is clearly visible and dominant at radii less than 0\farcs5. The rotation of the outer disc ($\gtrsim$0\farcs5) is very regular and relaxed. The dashed line shows the systemic velocity $V=4925\pm4$\,\kms.}
	\label{fig:PVD}
\end{figure}

A regularly rotating and symmetric molecular gas disc is clearly detected, with no evidence that the disc is disturbed by the strong AGN jets. 
It extends $\approx4$\arcsec$\times\,6$\arcsec\ in projection ($\approx1400\,\times\,1600$\,pc$^{2}$).
There is a slight dip in flux at the centre of the zeroth moment, partially due to our masking procedure removing low surface brightness emission spread over a large number of channels close to the central SMBH. This hole becomes much less significant when a simple clipping procedure is used, although this does increase the noise.
The enhanced velocities around the centrally-located SMBH are obvious in both the first moment map and the major-axis position velocity diagram (PVD; \reffig{fig:PVD}), the latter constructed by summing pixels within a 5-pixel wide (0\farcs175) pseudo-slit at a position angle of 142\degr. 
{The position angle used here and derived from the CO observations is reasonably consistent with the optical position angle as listed in the NASA/IPAC Extragalactic Database\,(NED)\footnote{\url{https://ned.ipac.caltech.edu}} of $\approx150^{\circ}$.
The moment one colourbar and PVD right-hand velocity axis are the observed line-of-slight velocity minus the systemic velocity of $V=4925\pm4$\,\kms\ derived in \refsec{sec:MCMC}. The large velocity dispersion observed at the centre of \reffig{fig:mom2} is primarily due to beam smearing.}

\begin{figure}
	\includegraphics[width=\columnwidth]{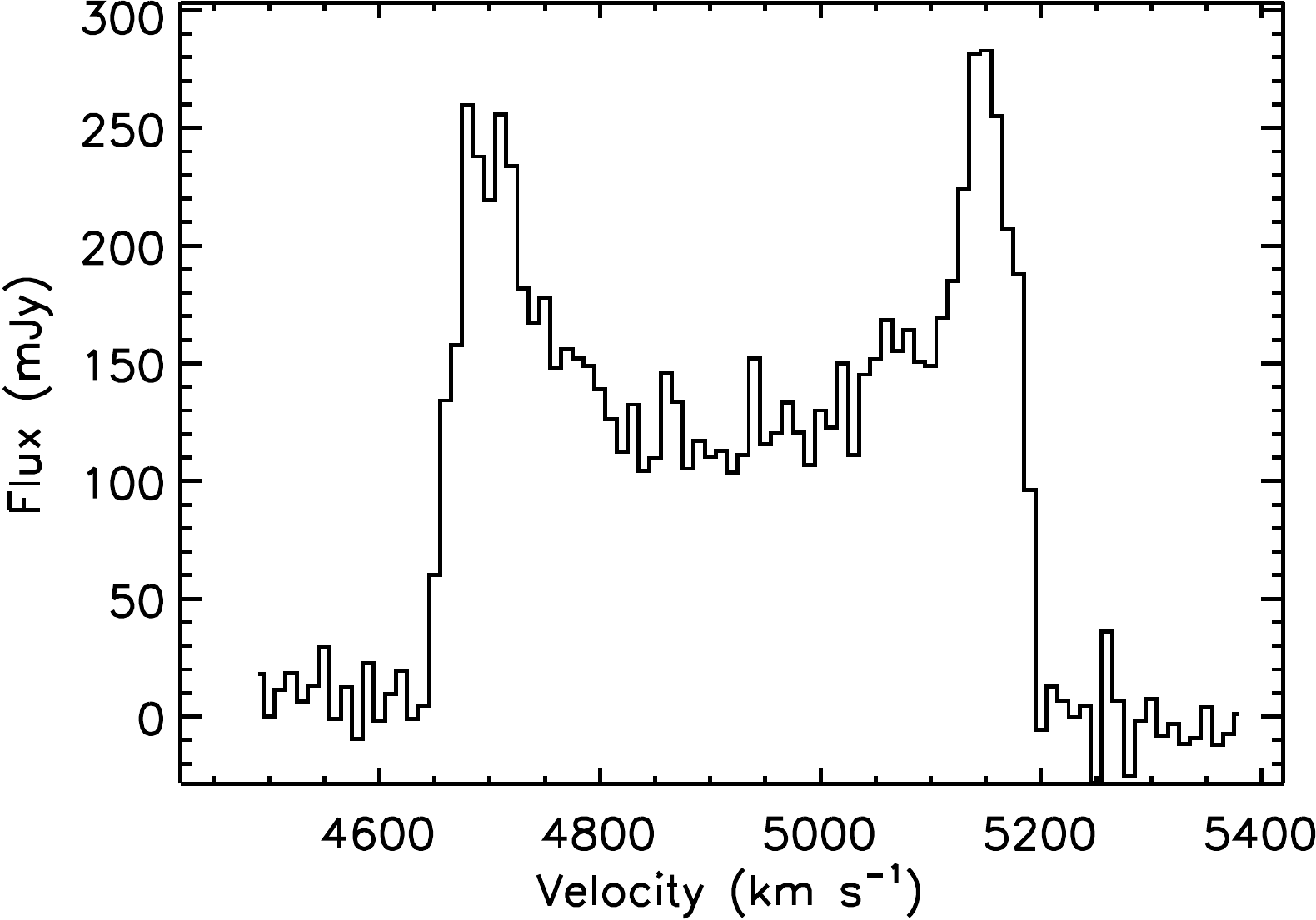}
	\caption{$^{12}$CO(2--1) integrated spectrum of NGC\,0383, showing the clear double-horned shape of a rotating disc.}
	\label{fig:spec}
\end{figure}



\reffig{fig:spec} shows the $^{12}$CO(2--1) integrated spectrum, made by integrating over a $6$\arcsec$\times6$\arcsec\ area of the clean cube, thus encompassing the entire disc. It clearly shows the double-horn shape of a rotating disc, {as also observed by \citet{Lim2000} in both CO(1--0) and CO(2--1) and \citet{Okuda2005} in CO(1--0) only. The total CO(2--1) flux is 87.1\,Jy\kms.}

For this work we want high spatial resolution data and hence use long-baseline interferometric observations. Incomplete $uv$--plane coverage can thus lead to some flux being resolved out. 
To check the scale of this problem, we compare the integrated flux derived from our CO(2--1) observations with that of \citet{OcanaFlaquer2010}, obtained with the 30-m IRAM single-dish telescope. Their CO(2--1) flux is $74.4\pm2.8$\,Jy\kms. As we retrieve slightly more flux than this, and the entire molecular gas disc of NGC\,0383 fits within the primary beam of the 30-m telescope, it is unlikely that we resolve out flux in our observations. The lower flux of the single-dish observations may be due to pointing and/or flux calibration errors. 

Comparing our CO(2--1) flux of 87.1\,Jy\kms\ to that of the CO(1--0) line (29.8\,Jy\kms; \citealt{OcanaFlaquer2010}), we find a CO(2--1)/CO(1--0) ratio of 0.73 after converting to beam temperature units (K\,\kms).
This ratio is very similar that found by \citet{Saintonge2017} in their mass-selected sample of local galaxies and within the range found by \citet{Leroy2013} for nearby star-forming disc galaxies, indicating the molecular gas in NGC\,0383 is similar to that in other local galaxies. 
The detection of CO line emission provides information about the cold gas mass distribution, that is later incorporated into our modelling (in addition to the kinematics themselves).

\subsection{Continuum Emission}
\begin{figure}
	\includegraphics[width=\columnwidth]{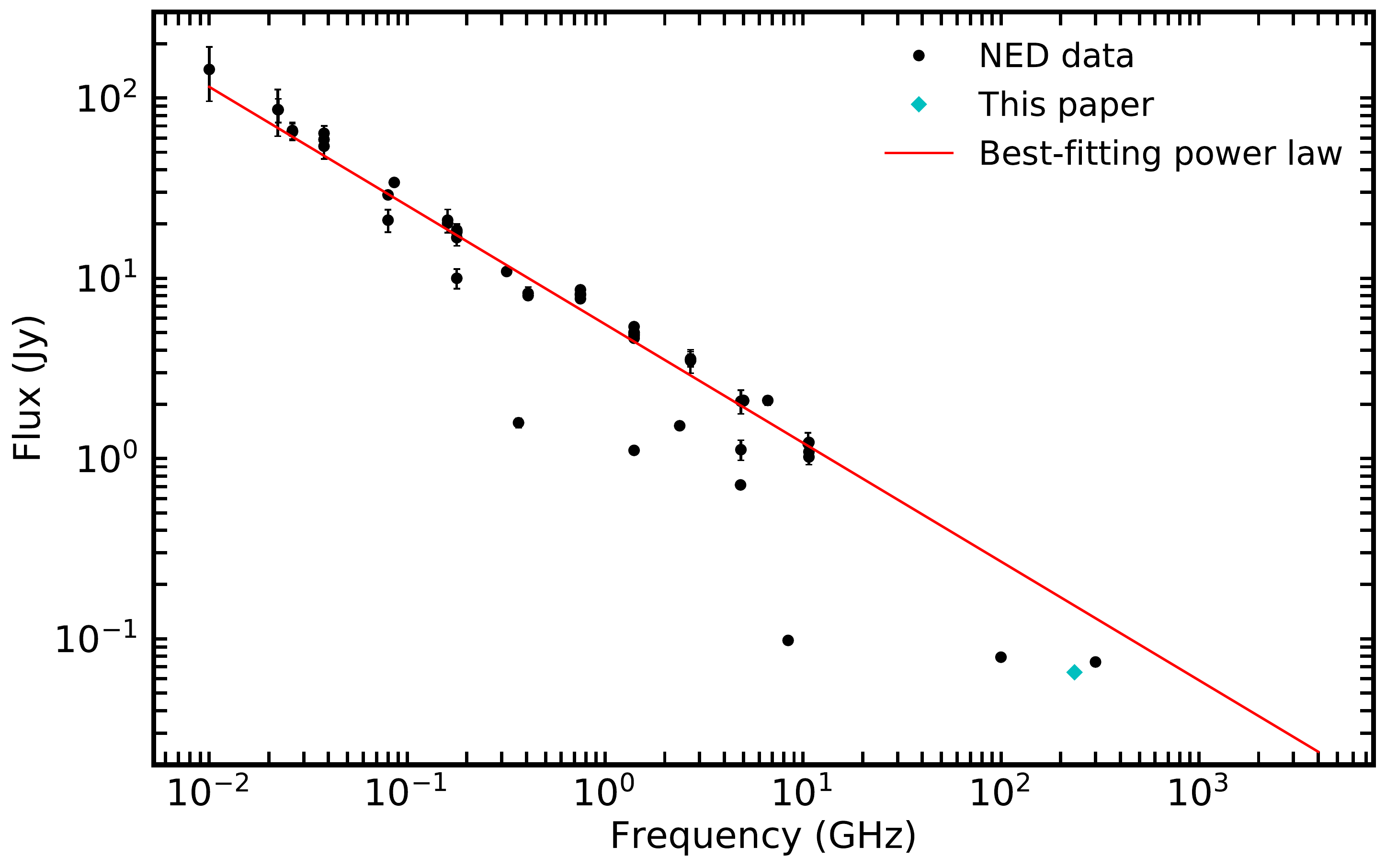}
	\caption{Spectral energy distribution of NGC\,0383 from radio to mm wavelengths, constructed using data from NED (black circles) and our data (cyan diamond). A best-fitting power law with slope -0.66 is overlaid in red. Error bars are plotted for all points but most are smaller than the symbol used.}
	\label{fig:SED}
\end{figure}

As mentioned previously, NGC\,0383 hosts a radio-loud AGN. We detect a continuum point source at the kinematic centre of the galaxy, with a total integrated intensity of 65.2$\pm$0.1\,mJy at a central frequency of 235.33\,GHz. Adding to our flux those tabulated in the NASA/IPAC Extragalactic Database\,(NED)\footnote{\url{https://ned.ipac.caltech.edu}} at millimetre and radio wavelengths, we constructed a radio--sub-mm spectral energy distribution (SED), shown in \reffig{fig:SED}. 
Our data point, shown by the cyan diamond, agrees well with previous observations. 
The literature data generally encompass emission from both the nucleus and the jets, but it is likely that it is the nucleus that causes the observed variability (i.e. the few data points well below the red  best-fitting line in \reffig{fig:SED}). 
Nevertheless, the data are fitted well with a simple power law for the flux $F$ as a function of frequency $\nu$ ($F_{\nu} \propto \nu^{\alpha}$), with a power-law index $\alpha = -0.66\pm0.03$ (the red line shown in \reffig{fig:SED}). 
This index value ($\approx$-0.7) is typical of a radio galaxy dominated by synchrotron radiation, as expected here from the prominent AGN jets (e.g. \citealt{Macdonald1968,Bridle1984,Laing2002}).

Despite the prominence of the AGN jets, the extreme regularity of the molecular gas distribution and kinematics (\reffig{fig:mom01}) indicates that the radio AGN activity does not directly disturb the gas disc. Our ability to model the disc motions and estimate the SMBH mass is thus unaffected. 

\section{Dynamical Modelling}
\label{sec:model}
The method we use to estimate the SMBH mass is described in detail in \citet{Davis2017} and was used in the previous WISDOM papers, but we summarise the specifics for modelling NGC\,0383 in this section. 
We make use of the publicly available \textsc{Kinematic Molecular Simulation}\,(KinMS)\footnote{\url{https://github.com/TimothyADavis/KinMS}} mm-wave observation simulation tool of \citet{Davis2013a} to create models of the data cube. 
KinMS uses input information about the gas distribution and kinematics, including a circular velocity curve.
Applying observational effects such as beam smearing and velocity binning, KinMS then creates a simulated data cube that can be directly compared to the observed data cube.
The model parameters are incrementally driven towards the best-fitting values by a Markov Chain Monte Carlo (MCMC) method. 
The MCMC algorithm fully samples the $\chi^2$ hyper-volume to estimate the posterior distributions and hence uncertainties on the best-fitting values. 


\subsection{Mass Model}
\label{sec:MGE}

We use an axisymmetric model of the stellar light distribution to derive the circular velocity curve of the galaxy.
We assume that the stellar mass dominates the potential in the inner parts of the galaxy: the molecular gas mass density is negligible in this system (see \refsec{sec:ml}), while dark matter is usually unimportant at small radii, as shown by e.g. \citet{Cappellari2013}. 
{Even if this latter assumption is incorrect, if the dark matter were distributed identically to the stellar mass in the inner parts of the galaxy, it would simply lead to a higher mass-to-light ratio and would not affect the best-fitting SMBH mass.
If dark matter were to contribute significantly and be distributed differently to the stellar mass, we would then find evidence for a significant mass-to-light ratio gradient (we find marginal evidence for a small mass-to-light ratio gradient in \refsec{sec:MCMC}).}

\begin{figure}
	\includegraphics[width=\columnwidth]{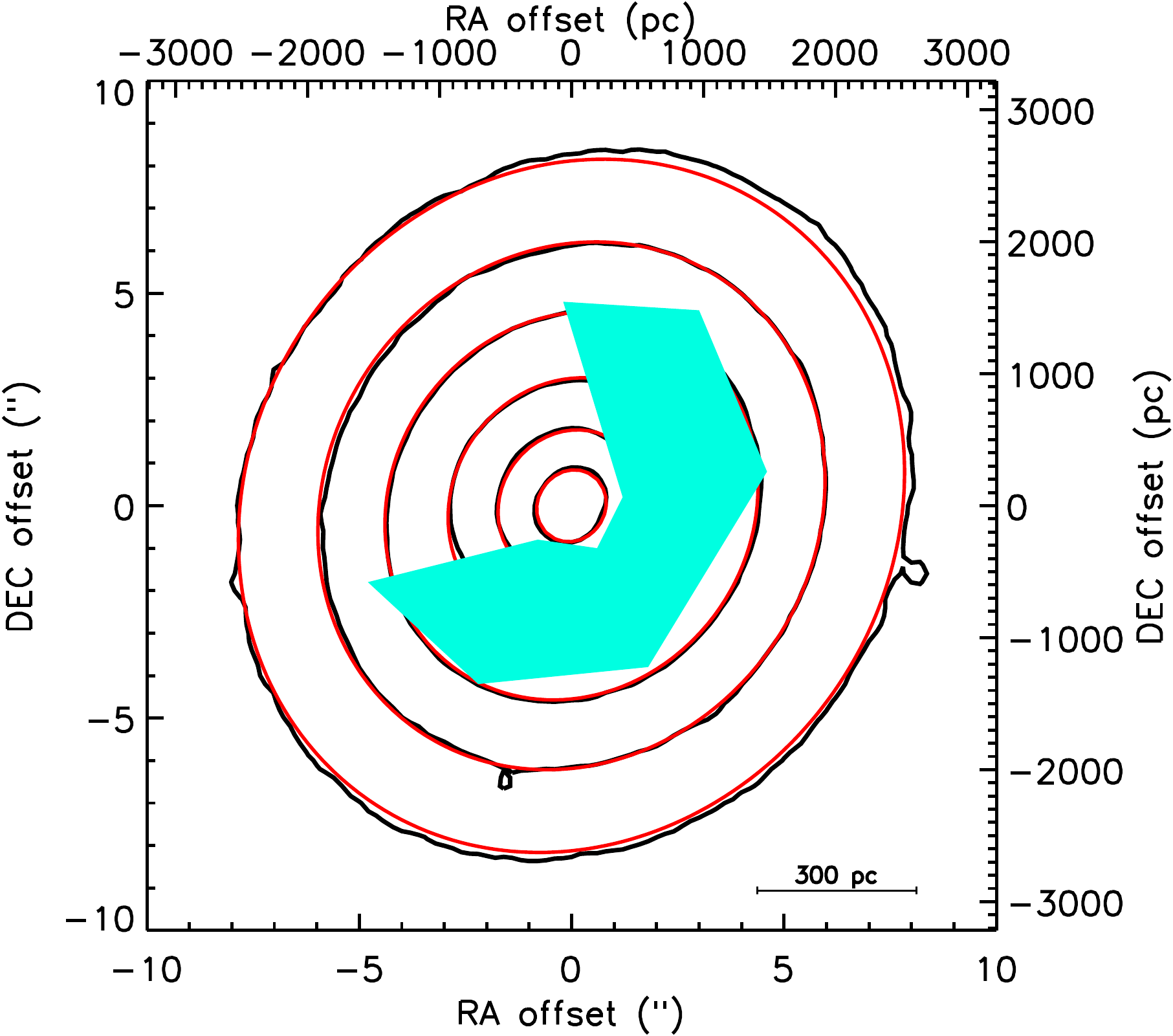}\hspace{0.5cm}
	\includegraphics[width=\columnwidth]{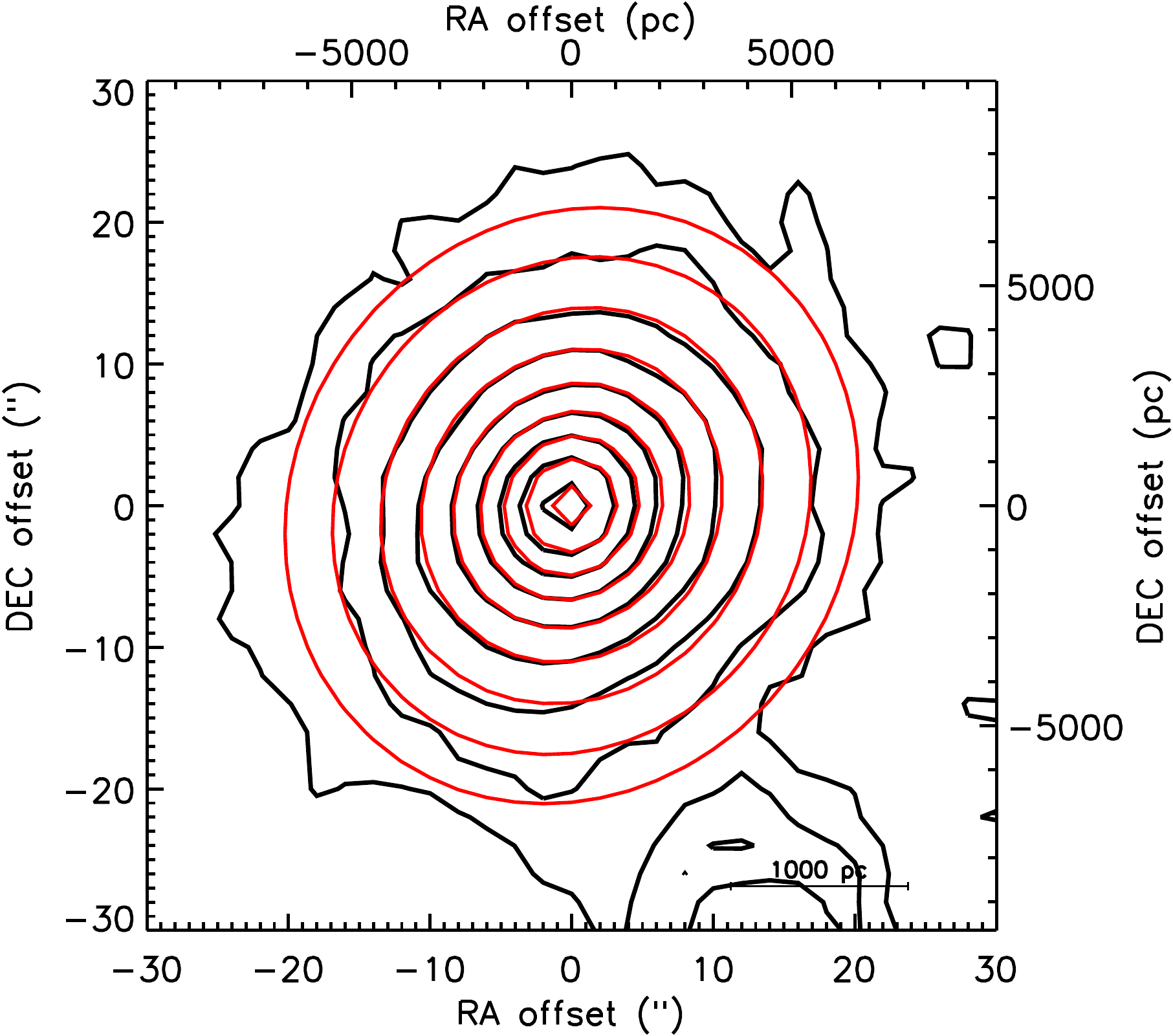}
	\caption{Multi-Gaussian Expansion (MGE) model of NGC\,0383 (red contours) overlaid on the \textit{HST} NICMOS F160W image (black contours, top panel) and the 2MASS $H$-band image (black contours, bottom panel). In the \textit{HST} image (top panel), the area masked due to dust is shown in cyan. A foreground star, bottom-right in the 2MASS image (bottom panel), is outside the fit radius and does not affect the MGE.}
	\label{fig:MGE}
\end{figure}

To model the luminous mass we perform a Multi-Gaussian Expansion (MGE; \citealt{Emsellem1994}), using the method implemented in the \textsc{MGE\_FIT\_SECTORS} Interactive Data Language (IDL) software\footnote{\label{fnt:jam}\url{http://purl.org/cappellari/software}, part of the Jeans Anisotropic MGE (JAM) dynamical modelling package of \citet{Cappellari2008}.} version v4.12 of \citet{Cappellari2002a}. 
We use a combined \textit{Hubble Space Telescope (HST)} Near Infrared Camera and Multi-Object Spectrometer (NICMOS) F160W and Two Micron All-Sky Survey (2MASS) $H$-band image.
This combined image allows us to model the stellar light with a sum of two-dimensional (2D) Gaussians up to a radius of 20\arcsec\,(6.4\,kpc), the \textit{HST} image being used exclusively for the inner $\approx$4\arcsec\ (1.3\,kpc) in radius because of its superior angular resolution.
To minimise the effect of dust attenuation on the mass-to-light ratio, the \textit{HST} image was masked over part of its lower-right limb (see the cyan region in \reffig{fig:MGE}, top panel).
The resulting MGE model is shown in \reffig{fig:MGE}, with the values of each Gaussian listed in \reftable{tab:MGE} (these values have not been deconvolved).

The circular velocity curve is then calculated by the \textsc{MGE\_CIRCULAR\_VELOCITY} procedure$^{\ref{fnt:jam}}$, by first analytically deprojecting the 2D Gaussians to a three-dimensional (3D) mass distribution, calculating the potential, and hence the circular velocity.
{The procedure uses a mass-to-light ratio of 1\msun/L$_{\odot,\,\mathrm{F160W}}$. 
The circular velocity is then multiplied element-wise by the square root of the mass-to-light ratio adopted, and a point mass representing the SMBH is added in the centre. The functional form of the mass-to-light ratio is fully explained in \refsec{sec:MCMC}.} 
However, we will show in \refsec{sec:discussion} that, in the case of NGC\,0383, the SMBH mass is essentially independent of the stellar mass-to-light ratio. 



\begin{table}
	\centering
	\caption{MGE best-fitting Gaussians (not deconvolved).}
	\label{tab:MGE}
	\begin{tabular}{ccc} 
		\hline
		I & $\sigma_{j}$ &  $q_{j}$ \\
		(L$_{\odot,\,\mathrm{F160W}}$ $\rm pc^{-2}$) & (\arcsec) & \\
		\hline
		12913.05 & \phantom{0}\phantom{0}\phantom{0}0.0682 & \phantom{0}0.91 \\
		\phantom{0}3996.80 & \phantom{0}\phantom{0}0.823 & 0.9 \\
		\phantom{0}5560.84 & \phantom{0}1.13 & \phantom{0}0.93 \\
		\phantom{0}4962.29 & \phantom{0}2.63 & 0.9 \\
		\phantom{0}2877.92 & \phantom{0}4.98 & 0.9 \\
		\phantom{0}\phantom{0}957.88 & 12.7\phantom{0} & 0.9 \\
		\hline
	\end{tabular}
\parbox[t]{0.8\columnwidth}{{{\textit{Notes:} For each Gaussian component, column 1 lists its F160W central surface brightness, column 2 its standard deviation (width) and column 3 its axis ratio.}}}
\end{table}

\subsection{Bayesian analysis}
\label{sec:MCMC}

We use a MCMC method to find the posterior distribution of the model best fitting the NGC\,0383 data, making use of the IDL \textsc{KinMS\_mcmc}\footnote{\url{https://github.com/TimothyADavis/KinMS\_MCMC}} code of \citet{Davis2013b, Davis2017}, that easily interfaces with the KinMS simulation tool to create new models and calculate and maximise the likelihood. 
A single fit was made to the whole clean data cube, with one hundred thousand iterations. 
All parameters had flat priors in linear space within specified physical limits, as listed in \reftable{tab:MCMC-params}, the only exception being the SMBH mass prior that was flat in log-space.
The observational errors were taken to be the RMS of the data cube in {line-free} channels, assumed to be constant throughout the cube.

The molecular gas disc of NGC\,0383 has a slight nuclear ring and outer spiral/ring structures that make assuming a smoothly-varying monotonic radial profile inappropriate.
Rather than construct an arbitrarily-complicated parametrisation of the radial gas distribution, we adopt instead the observed gas distribution as an input to our \textsc{KinMS} model.
Using the \textsc{SkySampler}\footnote{\url{https://github.com/Mark-D-Smith/KinMS-skySampler}} tool \citep{Smith2019}, we thus sample the de-convolved CLEAN components produced by the \textsc{CASA} task to generate a set of gas particles that exactly replicate the surface brightness profile. These particles are then used as an input into \textsc{KinMS}, with the three-dimensional central position, inclination and position angle of the gas disc as free parameters. The centre is initially assumed to be at the centre of the continuum emission (RA$=01^{\mathrm{h}}07^{\mathrm{m}}24.^{\mathrm{\!\!\!s}}95$, Dec.$=+32^{\circ}24\arcmin45\farcs15$) and the velocity of the central channel of the cube ($V_{\mathrm{helio,\,radio}}=4940$\,\kms).
{With no evidence to the contrary, we use the thin disc approximation for NGC\,0383.}

{We found that allowing a linearly-varying radial mass-to-light ratio profile fits the data better than a single (constant) mass-to-light ratio. Initial fits used a single mass-to-light ratio for the whole disc, but this did not provide a good fit to the entire data cube. 
We therefore implemented the simplest model to account for this, a linearly varying mass-to-light ratio, defined as 
\begin{equation}
M/L(R) = (M/L_{\mathrm{outer}}-M/L_{\mathrm{inner}}) \left(\frac{R}{3\farcs5}\right) + M/L_{\mathrm{inner}}\,,
\end{equation}
where $R$ is the radius and the inner ($M/L_{\mathrm{inner}}$) and outer ($M/L_{\mathrm{outer}}$) mass-to-light ratios are free parameters of our fit. The inner value is set at the centre of the disc ($R=0$\arcsec) with the outer edge at $R=3\farcs5$ and a flat mass-to-light ratio beyond that.}

Here we adopt the usual definition of $1\sigma$ ($3\sigma$) uncertainties as the 68.3\% (99.7\%) confidence intervals of the Bayesian posteriors found from the MCMC.
\reftable{tab:MCMC-params} lists the best-fitting value of each model parameter, along with its formal uncertainties.

\begin{figure*}
	\includegraphics[width=\textwidth]{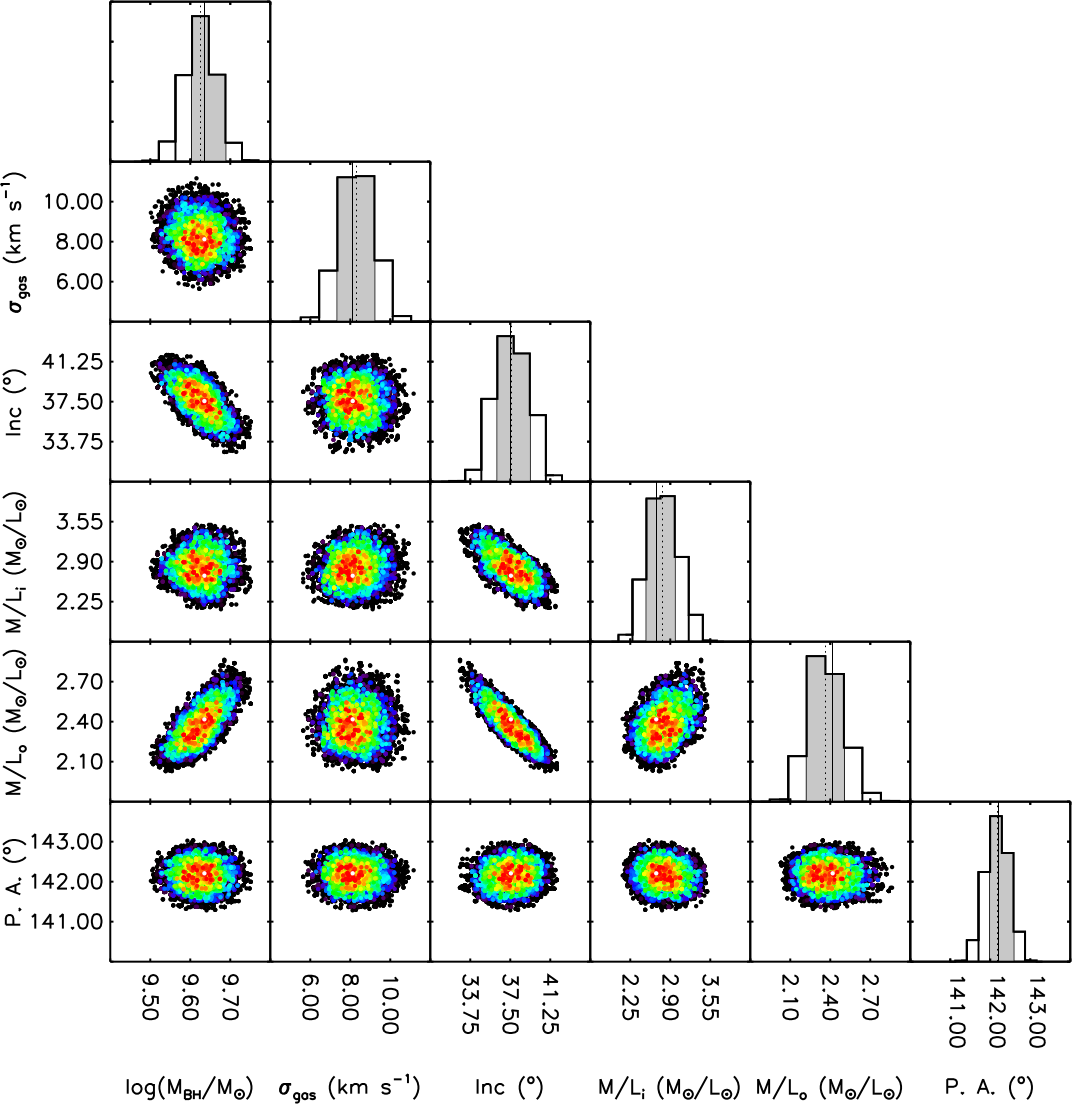}
	\caption{Corner plots showing covariances between the model parameters, for the non-nuisance parameters. The colours represent increasing confidence intervals from 68.3\% (red, $1\sigma$) to 99.7\% (blue, $3\sigma$). The white dots show the $\chi^{2}_{min}$ values. Covariances are present between the SMBH mass and outer stellar mass-to-light ratio, SMBH mass and inclination, and inclination and both stellar mass-to-light ratios. In the SMBH mass cases, this is exaggerated by plotting linear against logarithmic scales. Histograms show the one-dimensional marginalised posterior distribution of each model parameter. The shaded regions indicate the 68\% (1$\sigma$) confidence intervals. The black dashed lines show the median values and the black solid lines the $\chi^{2}_{min}$ values.}
	\label{fig:llconts}
\end{figure*}

As discussed in Section 3.2 of \citet{VanDenBosch2009}, when working with very large datasets the statistical uncertainties can be severely underestimated due to the dominance of the systematic uncertainties.
Accordingly, they suggest an approximate correction to account for the systematic uncertainties, by rescaling the $\Delta\,\chi^2$ (with respect to the minimum $\chi^2$, $\chi^{2}_{\mathrm{min}}$) required to define a given confidence level by the standard deviation of the $\chi^2$, namely $\sqrt{2(N-P)}\,\approx\,\sqrt{2N}$, where $N$ is the number of constraints ($\approx5.9\times10^{6}$) and $P$ is the number of inferred model parameters (10).
This sets the 68.3\% (99.7\%) confidence level at $\chi^{2}_{\mathrm{min}} +\sqrt{2N}$ ($\chi^{2}_{\mathrm{min}} +3\sqrt{2N}$). 
Applying this rescaling results in significantly larger uncertainties on the fitted parameters, which are likely to be more physically plausible. The same method was applied and discussed in detail by \citet{Smith2019}, and we use it here in the MCMC fitting of NGC\,0383. The corner plots and one-dimensional marginalisation of each model parameter are shown in \reffig{fig:llconts}.

Correlations are induced between pixels due to the synthesised beam, that can be corrected for by accounting for the induced covariance. However, the effect of this covariance on the MCMC uncertainties is negligible compared to the rescaling of the $\chi^{2}$ discussed above, hence we did not include the covariance matrix in our calculations.

\begin{figure*}
	\includegraphics[height=4.6cm,trim=0cm 0cm 2.3cm 0cm ,clip]{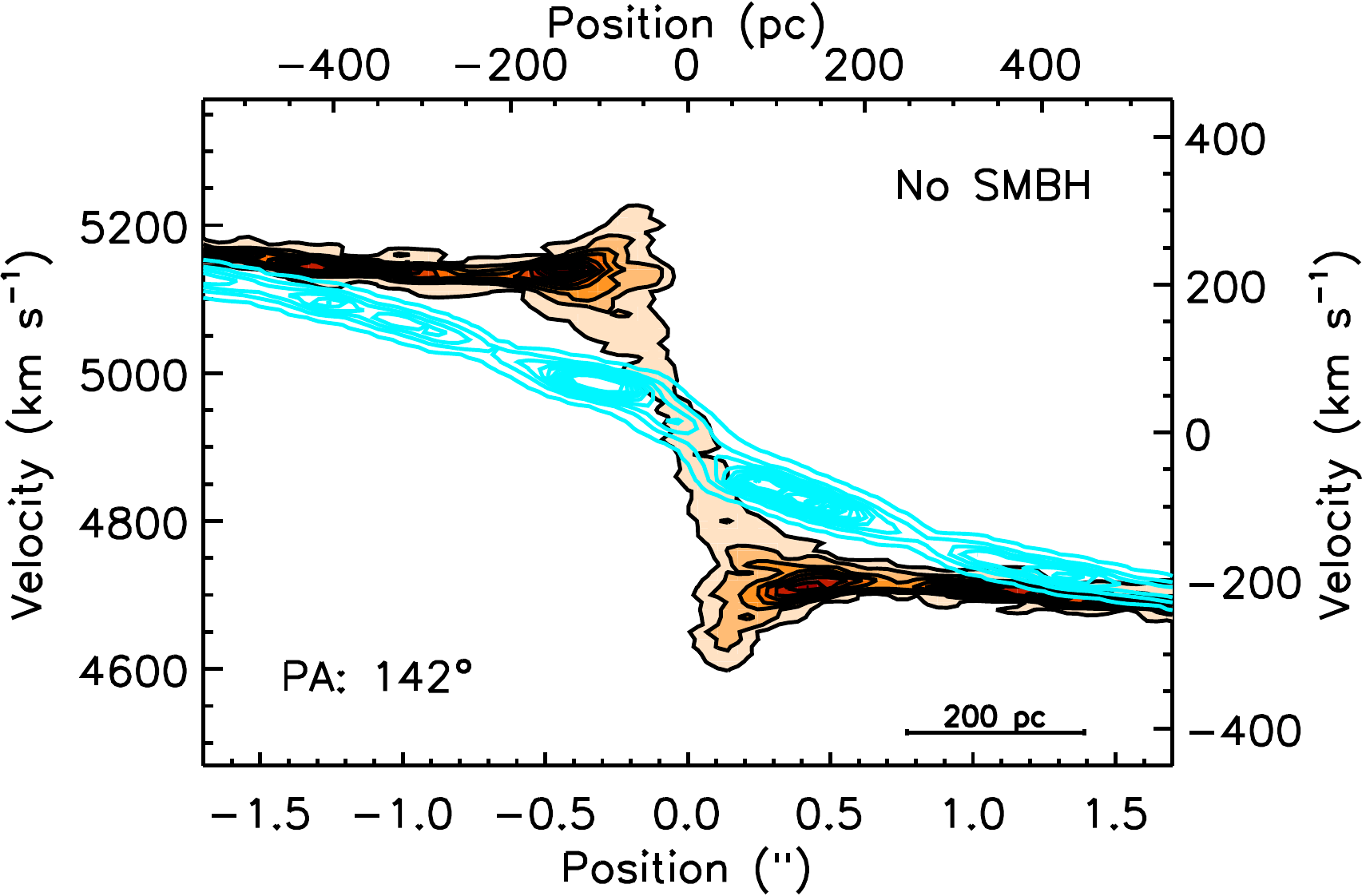}
	\includegraphics[height=4.6cm,trim=2.4cm 0cm 2.3cm 0cm ,clip]{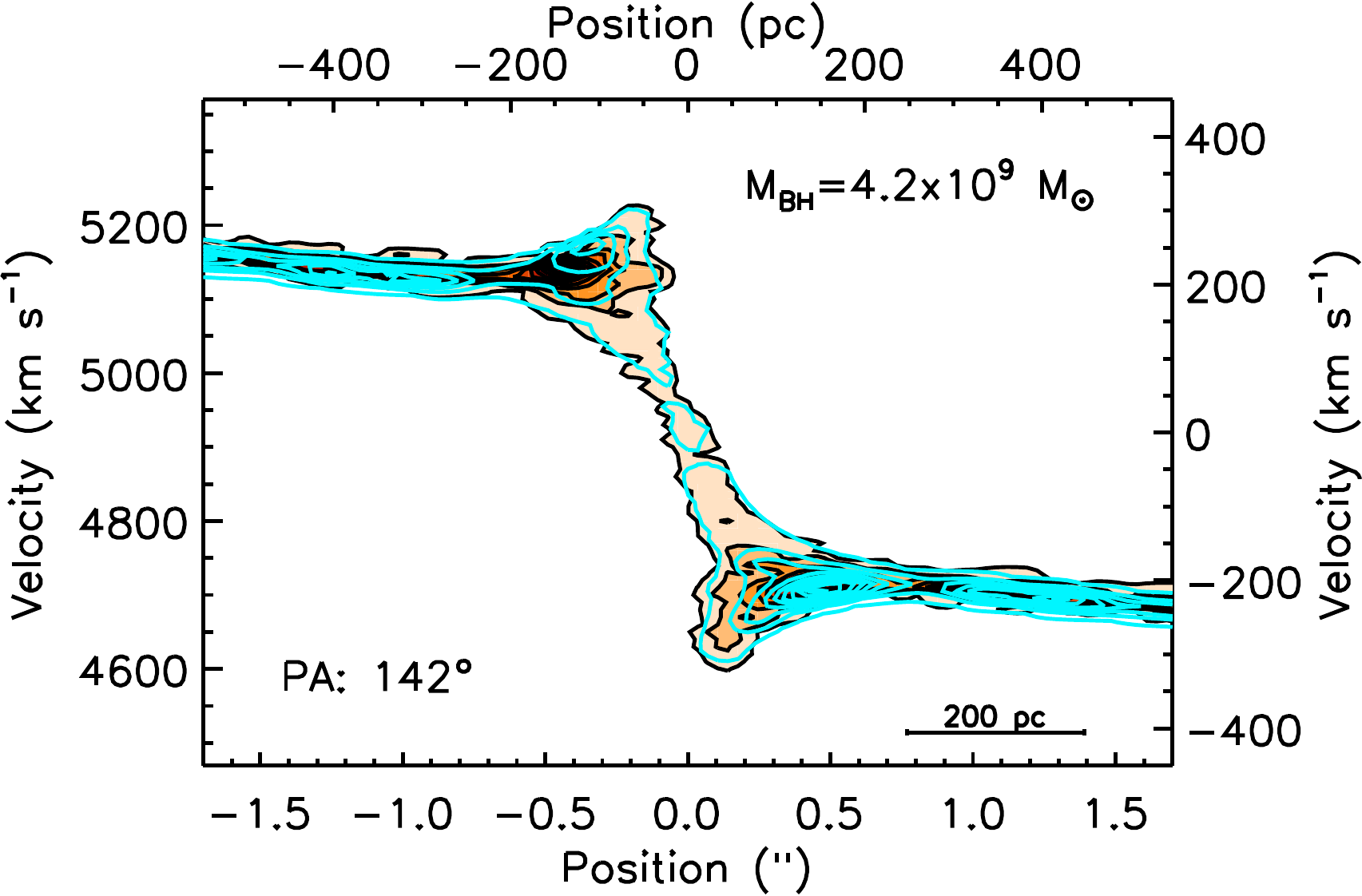}
	\includegraphics[height=4.6cm,trim=2.4cm 0cm 0cm 0cm ,clip]{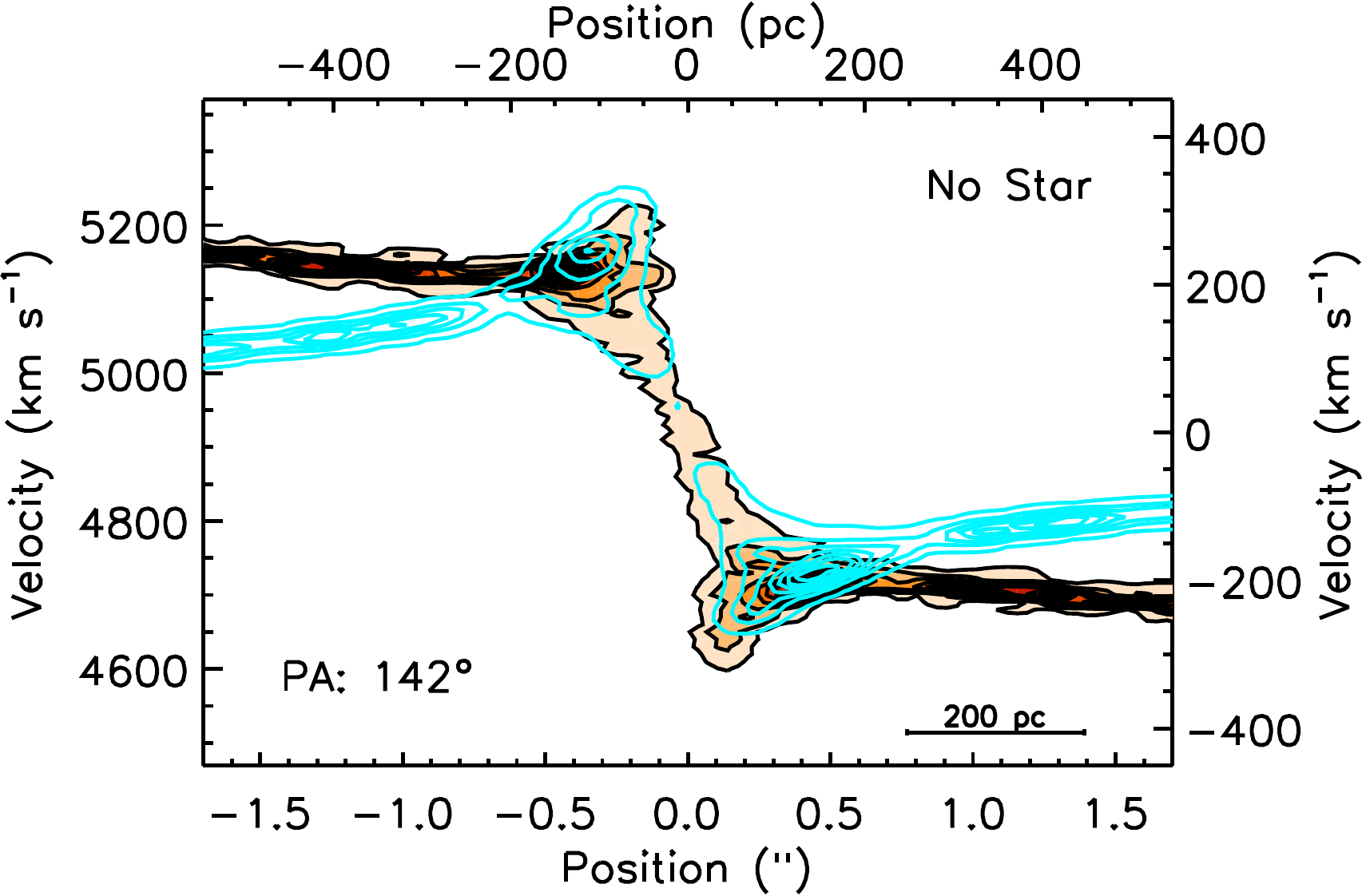}
	\caption{Observed position-velocity diagram (PVD) of NGC\,0383 with the smooth mask applied (orange contours) with the best-fitting model's PVD overplotted (blue contours; middle panel). The left panel shows the same model, but with no SMBH. The right panel also shows the same model, but with both mass-to-light ratios set to zero (no stellar contribution).}
	\label{fig:ModelPVD}
\end{figure*}

We find strong evidence for a SMBH, of mass (4.2$\pm$0.7)$\times$10$^{9}$\msun\ ($3\sigma$ uncertainty). The best-fitting model's PVD is shown in the middle panel of \reffig{fig:ModelPVD} as the blue contours overlaid on the data. It has a reduced $\chi^{2}$ of 1.01. 
Figure \ref{fig:ModelPVD} shows that a kinematic model with a dark massive object at the centre is the only model to fully describe the data.
In the left panel the SMBH has been removed and the model no longer reproduces the data. 
The right panel of \reffig{fig:ModelPVD} shows the best-fitting model with the mass-to-light ratio set to zero, i.e. no stellar mass, demonstrating the that SMBH mass dominates in the inner 0\farcs5 (as the fit is still very good in that region). 

{The best-fitting F160W-band mass-to-light ratio decreases linearly from 2.8$\pm$0.6\msun/\,L$_{\odot,\,\mathrm{F160W}}$ in the centre to 2.4$\pm$0.3\msun/\,L$_{\odot,\,\mathrm{F160W}}$ at the outer edge of the disc (both $3\sigma$ uncertainties). The spatial centre (as indicated by the X and Y offsets) is consistent with the unresolved continuum source to within the beamsize.}

\begin{table*}
	\begin{center}
	\caption{Best-fitting parameters with uncertainties from the MCMC fits.}
	\label{tab:MCMC-params}
	\begin{tabular*}{0.8\textwidth}{@{\extracolsep{\fill}}lcccc} 
		\hline
		Parameter & Search range & Best fit & $1\sigma$ uncertainty & $3\sigma$ uncertainty \\
		\hline
		SMBH mass (log(\msun)) & \phantom{0}\phantom{0}8.7\phantomsection{0}--9.95 & \phantom{0}9.63 & 0.04 & 0.08 \\
		Stellar $M/L$ inner (\msun/\,L$_{\odot,\,\mathrm{F160W}}$) & 0.01--10 & \phantom{0}2.78 & 0.21 & 0.61 \\ 
		Stellar $M/L$ outer (\msun/\,L$_{\odot,\,\mathrm{F160W}}$) & 0.01--10 & \phantom{0}2.36 & 0.12 & 0.33 \\
		\hline
		Position angle ($^\circ$) & \phantom{0}112--172 & 142.20\phantom{0} & 0.04 & 0.10\\ 
		Inclination ($^\circ$) & \phantom{0}26--89 & 37.58 & 1.67 & 3.48 \\ 
		Velocity dispersion (km s$^{-1}$) & \phantom{0}\phantom{0}0--15 & \phantom{0}8.32 & 0.72 & 2.11 \\ 
		\hline
		Nuisance parameters & & & & \\
		Integrated intensity (Jy\kms) & \phantom{0}\phantom{0}\phantom{0}5--200 & \phantom{0}74.60 & 4.15 & 9.79 \\
		Centre X offset (\arcsec) & -5--5 & \phantom{0}-0.00 & 0.01 & 0.03 \\
		Centre Y offset (\arcsec) & -5--5 & \phantom{0}-0.05 & 0.02 & 0.04 \\
		Centre velocity offset (km s$^{-1}$) & -50--10 & -15.16 & 1.37 & 3.63 \\
		\hline
	\end{tabular*}
\parbox[t]{0.8\textwidth}{{\textit{Note:} The X and Y offsets are measured with respect to the location of the unresolved continuum point source,  RA$=01^{\mathrm{h}}07^{\mathrm{m}}24.^{\!\!\!\mathrm{\!s}}96$ and Dec $=+32^{\circ}24'45\farcs11$. The velocity offset is measured with respect to the central channel of the cube ($V_{\mathrm{helio, radio}}=4940$\,\kms). The best-fitting centre velocity offset thus defines a systemic velocity of $V=4925\pm4$\,\kms.}}
\end{center}
\end{table*}

\section{Discussion}
\label{sec:discussion}

In this work we have presented ALMA $^{12}$CO(2--1) observations of NGC\,0383 showing a relaxed gas disc (\refsec{sec:Obs}). The data clearly show the kinematic signature of a massive dark object, with a mass of ($4.2\pm0.7$)$\times10^{9}$\,\msun\ ($3\sigma$ uncertainty) measured through dynamical modelling (\refsec{sec:model}). 

\subsection{Uncertainties}
The uncertainties associated with a SMBH mass derived through the molecular gas technique are discussed extensively in the previous papers using this method. Each paper builds from the last and focuses on the sources of uncertainty that are relevant to each galaxy. In particular, \citet{Smith2019} discussed properly constraining the mass-to-light ratio and inclination. As we resolve the SMBH $R_{\rm SOI}$ in NGC\,0383 in this case our mass measurement is essentially independent of the mass-to-light ratio (see Section \ref{sec:ml}). Accurately determining the inclination, however, remains important. Our choice to apply the $\chi^2$ scaling (discussed in Section \ref{sec:MCMC}) allows us to retrieve more physically meaningful estimates of our inclination uncertainties. As NGC\,0383 is fairly face-on ($i\approx38^{\circ}$) these uncertainties dominate the error budget, through the degeneracy between inclination and SMBH mass (see Figure \ref{fig:llconts}). 

\begin{figure}
	\includegraphics[width=\columnwidth]{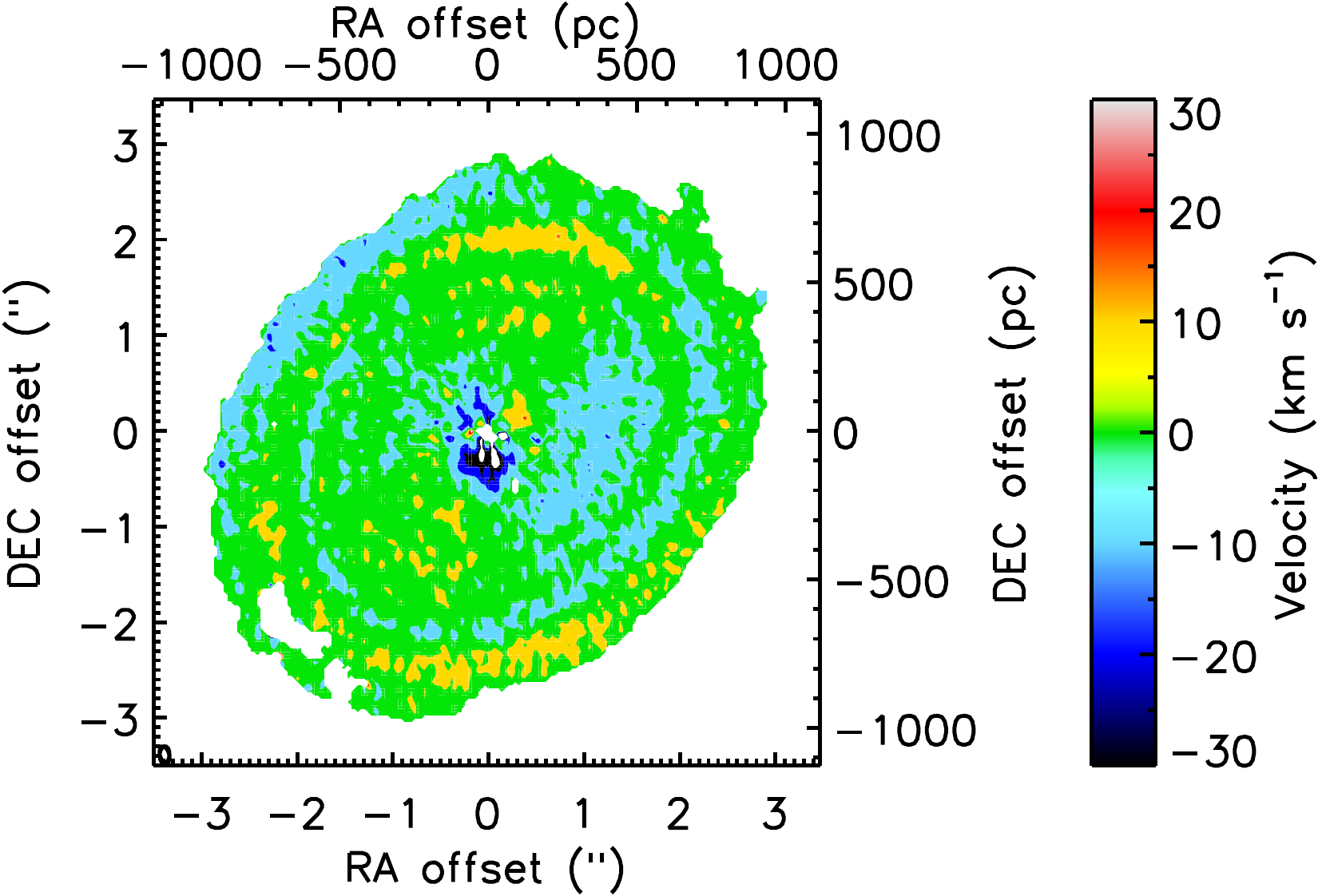}
	\caption{First moment (intensity-weighted mean velocity) residuals of NGC\,0383, created by subtracting the first moment of the best-fitting model cube from the first moment of the data cube. The plot clearly shows the slight spiral features that could not be modelled by our axisymmetric mass model. Due to the simplicity of the model and weighting when creating the first moment, there are larger velocity residuals near the disc centre.}
	\label{fig:resid}
\end{figure}

Other potential sources of uncertainty arise from the assumption that the molecular gas is dynamically cold and rotating on circular orbits. The velocity dispersion of the gas is consistently small ($\sigma_{\mathrm{gas}}<\,10$\kms), indicating the disc is nearly perfectly rotationally supported ($V_{\mathrm{rot}}/\sigma_{\mathrm{gas}}\,\gtrsim\,40$, where $V_{\mathrm{rot}}$ is the deprojected rotation velocity of the gas in the nearly flat portion of the rotation curve; see e.g. \reffig{fig:PVD}).
Despite this, some non-circular motions do appear to be present. The velocity residuals (\reffig{fig:resid}; data moment 1 minus best-fitting model moment 1) show the same spiral/ring structures noted in the moment zero (\reffig{fig:mom01}), indicating that material may be flowing along these arms (potential fuelling the AGN). The larger velocity residuals near the centre are due to the intensity weighting when creating the moment 1 map.
However, the dominance of the SMBH in the central regions (see \reffig{fig:massprof}) indicates that non-circular motions are unlikely to significantly affect the derived SMBH mass.

All these uncertainties are small, and in any case they are dwarfed by that on the distance measurement. This uncertainty is $\approx15\,$\% (i.e. $66.6\pm9.9$\,Mpc), from the use of the Tully-Fisher relation in \citet{Freedman2001} to estimate the distance. The SMBH mass measurement scales linearly with the distance adopted, and as is customary we do not include the distance uncertainty in our result.

\subsection{Mass-to-light Ratio Influence}
\label{sec:ml}

The posterior distribution between SMBH mass and mass-to-light ratio shows a strong covariance (see \reffig{fig:llconts}, middle panels of the leftmost column), although this is exaggerated by plotting linear against logarithmic scales. The correlation present is contrary to the expected anti-correlation, and it may be a product of the SMBH mass--inclination and mass-to-light ratio--inclination correlations. By allowing the inclination to vary during the fit, the correlation between mass-to-light ratio and inclination dominates and induces correlations in other variables. See \citet{Smith2019} for a fuller discussion of this issue.  

\begin{figure}
	\includegraphics[width=\columnwidth]{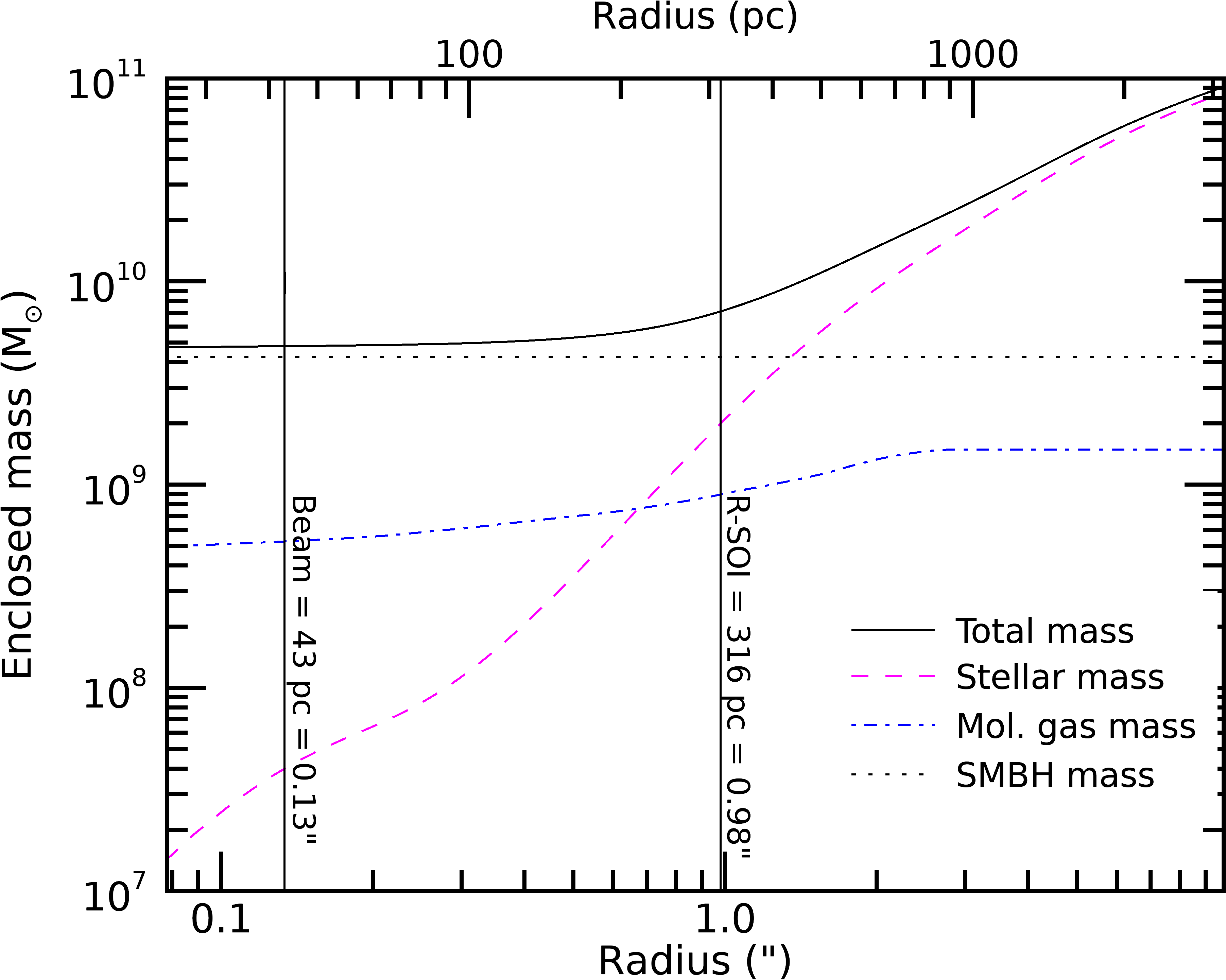}
	\caption{Cumulative mass profile of NGC\,0383, showing the total mass (black solid line), SMBH mass (black dotted line), stellar mass (magenta dashed line) and molecular gas disc mass (blue dot-dashed line) enclosed as a function of radius. The vertical lines indicate the synthesised beam and measured $R_{\rm SOI}$. At a radius of one synthesised beam, both the stellar mass and the molecular gas mass are insignificant ($\approx1\,\%$ and $\approx10\,\%$ of the total mass, respectively).}
	\label{fig:massprof}
\end{figure}

A simple calculation of the total mass enclosed from the circular velocity (and assuming spherical symmetry, i.e. $M_{<R} \propto V^{2}_{\mathrm{rot}}(R)/R$) allows us to determine how significant the stellar mass is as a function of radius in NGC\,0383. 
Figure \ref{fig:massprof} shows the enclosed stellar mass as a function of radius as well as the enclosed total mass, revealing that the stellar mass becomes significant only at a radius of $\approx$300\,pc. 
At 0\farcs13 (i.e. one synthesised beam, 43\,pc), the stellar mass is only $\approx1\,\%$ of the mass enclosed at that radius, so is insignificant. The molecular gas disc mass at this radius is $\approx10\,\%$ of the total enclosed mass, that is again small compared to the SMBH mass ($\approx90\,\%$ of the total enclosed mass).
This indicates that whilst the mass-to-light ratio (and assumed $X_{\rm CO}$) do have a covariance with the SMBH, their effect on the best-fitting value is very small and the SMBH mass is largely independent of them. 
The fact that our SMBH mass is almost independent of our luminous mass model in turn leads to the very small uncertainties on $M_{\rm BH}$ compared to other works (indeed, the $3\sigma$ confidence interval in \reffig{fig:llconts} is very narrow). It also gives us greater confidence in our measurement.

{Although in NGC\,0383 the SMBH dominates the total mass distribution within a few synthesised beams (i.e. angular resolution elements), this might not always be the case and in some instances the mass of the molecular gas disc itself may matter. In fact, even in NGC\,0383, the stellar and molecular gas masses are approximately equal at $R_{\rm SOI}$. This reinforces the importance of spatially-resolved molecular gas data for SMBH mass measurements.}

\subsection{Estimating $M_{\mathrm{BH}}$ from the Observed $R_{\mathrm{SOI}}$}
In contrast to the detailed dynamical modelling of \refsec{sec:MCMC}, we can make a crude estimate of the SMBH mass from the observed radius of the SMBH sphere of influence ($R_{\mathrm{SOI}}$).  
The SMBH $R_{\mathrm{SOI}}$ is defined as the radius within which the SMBH dominates the potential (see \refeq{eq:SOI}).
This radius can be determined from the observed PVD (\reffig{fig:PVD}) as the local minimum in the rotation curve (i.e the transition point) between the SMBH-dominated Keplerian curve ($V_{\mathrm{rot}} \propto 1/\sqrt{R} $) and the stellar mass-dominated approximately flat rotation curve ($V_{\mathrm{rot}} \simeq\,$constant).
{By visual inspection, we estimate this occurs at a radius of $\approx0\farcs7$ (see \reffig{fig:ModelPVD}). Using \refeq{eq:SOI} and $\sigma_{\ast}\,=\,239\pm16$\kms \citep{Bosch2016} 
then yields $M_{\rm BH} \approx 3.0\times$10$^{9}$\msun. 
Given that the SMBH $R_{\rm SOI}$ is so well resolved, this back-of-the-envelope estimate agrees well with our full modelling value of (4.2$\pm$0.7)$\times$10$^{9}$\msun\ (the latter also yielding an exact $R_{\rm SOI}=316\pm60$\,pc or $0\farcs98\pm0\farcs18$).}

\subsection{Comparison to the Literature}
\label{sec:lit}
An upper limit on the SMBH mass in NGC\,0383 has previously been determined by \citet{Beifiori2009}.
Once scaled to our distance (66.6\,Mpc from 63.4\,Mpc) and inclination (37\fdg6 from 33\degr) this is $M_{\rm BH} = 1.1\times10^{9}$\,\msun.
Given that this is lower than our measurement, it might indicate the presence of very disturbed ionised gas. 

\begin{figure}
	\includegraphics[width=\columnwidth]{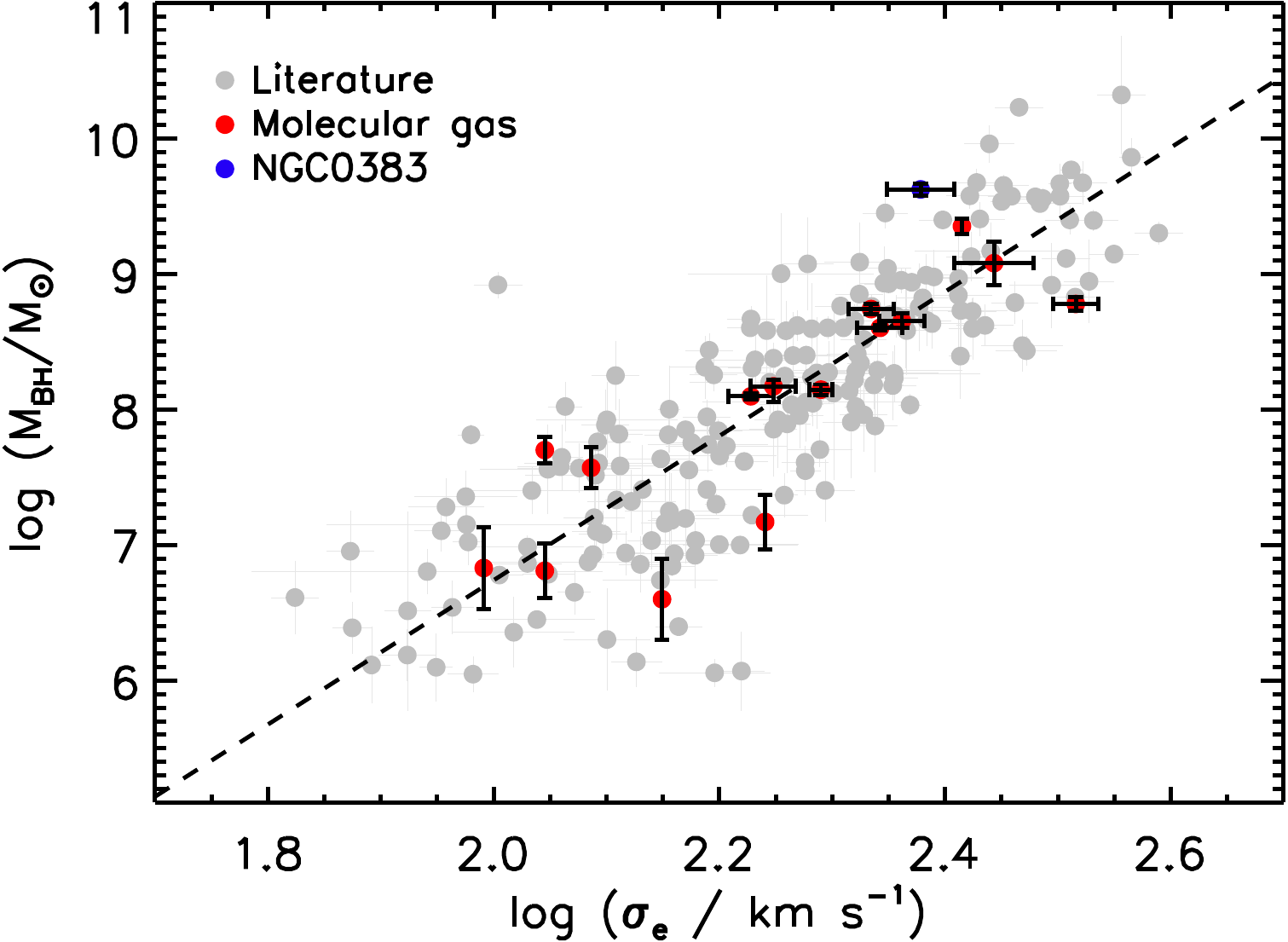}
	\caption{The $M_{\mathrm{BH}}-\sigma_{\rm e}$ relation from literature measurements (grey points and dotted line), as compiled by \citet{Bosch2016}. SMBH mass measurements using the molecular gas method are highlighted in red, while this work (NGC\,0383) is in blue. For the molecular gas-derived SMBH masses, the error bars shown correspond to $1\sigma$ uncertainties.}
	\label{fig:msig}
\end{figure}

One of the tightest known correlations between SMBH mass and a host galaxy property is that with the stellar velocity dispersion, i.e. the $M_{\rm BH}-\sigma_{\ast}$ relation \citep[e.g.][]{Gebhardt2000,Ferrarese2000}. We added our measurement to the dynamical measurements and power-law fit of \citet{Bosch2016} in \reffig{fig:msig}, to see whether our measurement also lies on this relation. The data of \citet{Bosch2016} are shown in grey, while our new measurement for NGC\,0383 is shown in blue. Other SMBH masses estimated using the molecular gas method are shown in red (\citealt{Davis2013b,Onishi2015,Barth2016a,Barth2016b,Onishi2017,Davis2017,Davis2018,Smith2019,Combes2019,Nagai2019,Boizelle2019}). 
NGC\,0383 has the largest SMBH mass estimated with molecular gas so far, and is on the upper edge of the scatter in the \citet{Bosch2016} $M_{\rm BH}$-$\sigma_{\ast}$ relation.
If accurate, the low $\sigma_{\mathrm{e}}$ compared to the SMBH mass indicates NGC\,0383 might be part of the so-called "over-massive black hole" population. These galaxies are thought to be local analogues of the higher-redshift quiescent galaxies that also contain over-massive black holes, and could therefore be relics that have evolved little since a redshift $z\approx2$ \citep{Bosch2016,Walsh2016}.
{Alternatively, an over-massive SMBH may be the result of merger(s), especially when paired with a high molecular gas mass in an ETG harbouring an AGN. \citet{Lim2000} discuss whether the AGN activity in NGC\,0383 is the result of a gas-rich minor merger, with black hole accretion beginning soon after the merger event. While this is plausable, it is unclear if this scenario could explain such an overly massive SMBH.}


\subsection{Comparison of Spatial Scales Probed by Molecular Gas and Megamasers}

Modelling megamaser dynamics is typically the most accurate method of measuring SMBH masses, due to the exquisite angular and spectral resolution usually achieved. The Keplerian rise we detect in the centre of NGC\,0383 indicates that our data reach very close to the SMBH.
The connection between the accretion disc/torus region (where masers are typically found) and the outer molecular gas disc has only recently begun to be explored in any detail. It is thought that position angle mismatches are common between these two components, {and tilted and counter-rotating accretion discs are frequently observed \citep[e.g. recently by][]{Imanishi2018,Combes2019}}. In contrast to this expectation, NGC\,0383 seems to have a single, unwarped molecular disc extending from kiloparsec scale to well within its SMBH SOI. 

We are able to estimate just how close to the SMBH the highest velocity molecular gas we detect here is. Equating the centrifugal and gravitational forces at a radius $R$ and assuming the SMBH mass dominates the stellar mass within this radius, we obtain
\begin{equation}
R = \frac{GM_{\mathrm{BH}}}{V^{2}_{\mathrm{c}}} ,
\end{equation}
where $V_{\rm c}$ is the circular velocity at $R$.
If we normalise the radius by the Schwarzschild radius
\begin{equation}
R_{\mathrm{Schw}} \equiv \frac{2GM_{\mathrm{BH}}}{c^{2}} ,
\end{equation}
where $c$ is the speed of light, and the circular velocity by $c$, we find
\begin{equation}
\frac{R}{R_{\mathrm{Schw}}} = \frac{1}{2}\left(\frac{V_{\mathrm{c}}}{c}\right)^{-2} ,
\end{equation}
where interestingly the SMBH mass has dropped out. All rotationally-supported discs around a SMBH should thus follow this unique relation, irrespective of the SMBH mass.
Substituting $V_{\mathrm{c}} = V_{\mathrm{obs}}/\sin(i)$, where $V_{\rm obs}$ is the observed line-of-sight velocity (along the galaxy major axis) and $i$ the inclination, we obtain
\begin{equation}
\frac{R}{R_{\mathrm{Schw}}} = 0.5\times10^{6} \left(\frac{300\,\mathrm{km\,s^{-1}} \sin(i)}{V_{\rm obs}}\right)^{2}.
\label{eq:RSchw}
\end{equation}

The maximum rotation velocity observed in NGC\,0383 is $V_{\rm obs}\,\approx\,350$\kms\ (the peak of the PVD in \reffig{fig:PVD}) and $i\,=\,37\fdg5$. The highest velocity molecular {gas} we detect therefore reaches $\approx\,1.36\times10^{5}$ Schwarzschild radii.

Megamasers, although rare, are the current gold standard for dynamical SMBH mass measurements. Megamasers are thought to trace gas very close to the SMBH (in the accretion disc/torus), and as such they probe the gravitational field of the SMBH in a way that is unaffected by most outside sources. In addition, in the best cases, maser observations provide independent geometric distance estimates, vastly reducing the dominant systematic effect that plagues most SMBH mass measurements. 
{Some of the earliest megamasers discovered were in NGC\,4258 \citep{Nakai1993,Herrnstein1999}, more recently The Megamaser Cosmology Project (MCP) have carried out the most complete survey of megamasers to date, with the goal of measuring Hubble's constant (see e.g. the survey compilation by \citealt{Braatz2015}). The MCP observations also allow them to make several SMBH mass measurements (e.g. \citealt{Reid2009}; \citealt{Zhao2018}).
The observed megamasers with SMBH masses have $V_{\rm obs}$ ranging from 170\kms\ \citep[NGC\,1029;][]{Gao2017} to 950\kms\ \citep[NGC\,2273;][]{Kuo2011}, with an average of $\approx600$\kms. 
All megamaser systems are observed close to edge-on, so that $\sin(i)\approx1$. Given this, megamasers typically probe gas at radii between $5\times10^{4}$ and $1.5\times10^{6}$ Schwarzschild radii.
Our data thus show that the molecular gas disc in NGC\,0383 extends unbroken and unwarped down to very close to the SMBH, and that it traces the same material probed by megamasers in other galaxies.}


\section{Conclusions}
\label{sec:conclusion}

We have presented a measurement of the mass of the SMBH in the nearby lenticular galaxy NGC\,0383 (radio source 3C\,031). This estimate is based on ALMA observations of the $^{12}$CO(2--1) emission line with a physical resolution of $\approx43$\,pc ($0\farcs18\times0\farcs1$). {We thus have a spatial resolution a factor of $>7$ better than the $R_{\rm SOI}$.} Our spectroscopic resolution, and a channel width of 10\kms, allow us to resolve gas down to $\approx140,000$ Schwarzschild radii and thus to probe the same material as typical megamaser observations.
NGC\,0383 has a relaxed, smooth nuclear disc with weak ring/spiral features. We detect a clear Keplerian increase of the rotation velocity of $^{12}$CO(2--1) at radii $\lesssim$0\farcs5, and forward model of our ALMA data cube with the KinMS tool in a Bayesian MCMC framework to measure a SMBH mass of ($4.2\pm0.7$)$\times10^{9}$\msun, a F160W-band mass-to-light ratio varying linearly from $2.8\pm0.6$\msun/\,L$_{\odot,\,\mathrm{F160W}}$ in the centre to $2.4\pm0.3$\msun/\,L$_{\odot,\,\mathrm{F160W}}$ at the outer edge of the molecular gas disc (3\farcs5 radius) and a velocity dispersion of $8.3\pm2$\kms\ (all $3\sigma$ uncertainties). We also detect continuum emission from the AGN in NGC\,0383 across the full bandwidth, consistent with synchrotron radiation.
This work not only shows the power of ALMA to estimate SMBH masses, but it also demonstrates that the molecular gas method is highly complimentary to megamaser observations as it can probe the same emitting material.

\section*{Acknowledgements}


EVN, MDS and TGW acknowledge support from a Science and Technology Facilities Council (STFC) PhD studentship.

TAD acknowledges support from a Science and Technology Facilities Council Ernest Rutherford Fellowship.
MB was supported by the consolidated grants `Astrophysics at Oxford' ST/H002456/1 and ST/K00106X/1 from the UK Research Council.
MC acknowledges support from a Royal Society University Research Fellowship.

The authors thank the referee for their comments that have improved this paper.

This paper makes use of the following ALMA data: ADS/JAO.ALMA\#2015.1.00419.S and ADS/JAO.ALMA\#2016.1.00437.S. ALMA is a partnership of ESO (representing its member states), NSF (USA) and NINS (Japan), together with NRC (Canada), NSC and ASIAA (Taiwan) and KASI (Republic of Korea), in cooperation with the Republic of Chile. The Joint ALMA Observatory is operated by ESO, AUI/NRAO and NAOJ.

This paper also makes use of observations made with the NASA/ESA Hubble Space Telescope, and obtained from the Hubble Legacy Archive, which is a collaboration between the Space Telescope Science Institute (STScI/NASA), the Space Telescope European Coordinating Facility (ST-ECF/ESA) and the Canadian Astronomy Data Centre (CADC/NRC/CSA).  
Data were also used from the Two Micron All-Sky Survey (2MASS) at IPAC. The 2MASS project is a collaboration between The University of Massachusetts and the Infrared Processing and Analysis Center (JPL/ Caltech). Funding is provided primarily by NASA and the NSF. The University of Massachusetts constructed and maintained the observatory facilities, and operated the survey. All data processing and data product generation is being carried out by IPAC. Survey operations began in Spring 1997 and concluded in Spring 2001.
This research has made use of the NASA/IPAC Extragalactic Database (NED) which is operated by the Jet Propulsion Laboratory, California Institute of Technology, under contract with the National Aeronautics and Space Administration.





\bibliographystyle{mnras}
\bibliography{NGC0383_v14Papers} 





\bsp	
\label{lastpage}
\end{document}